\documentclass[letterpaper]{article} %
\usepackage{aaai2026-underreview}  %
\usepackage{times}  %
\usepackage{helvet}  %
\usepackage{courier}  %
\usepackage[hyphens]{url}  %
\usepackage{graphicx} %
\usepackage{natbib}  %
\usepackage{caption} %
\usepackage[hidelinks]{hyperref}
\usepackage[nameinlink,noabbrev]{cleveref} %
\usepackage{tikz}
\usetikzlibrary{positioning, fit, backgrounds}
\usepackage{enumitem}
\usepackage{booktabs}
\usepackage{colortbl} %
\definecolor{level0}{RGB}{255,255,255}  %
\definecolor{level1}{RGB}{213,233,213}  %
\definecolor{level3}{RGB}{114,184,114}  %

\usepackage[title]{appendix}
\usepackage{amssymb}
\newcommand{\checkbox}[1]{%
  \item[$\square$] #1
}

\title{Audit Cards: Contextualizing AI Evaluations}

\author{
    Leon Staufer\textsuperscript{\rm 1,2}\equalcontrib,
    Mick Yang\textsuperscript{\rm 1,3}\equalcontrib,
    Anka Reuel\textsuperscript{\rm 4},
    Stephen Casper\textsuperscript{\rm 5,1}
}
\affiliations{
    \textsuperscript{\rm 1}ML Alignment \& Theory Scholars \\
     \textsuperscript{\rm 2}Technical University of Munich\\
    \textsuperscript{\rm 3}University of Pennsylvania\\
    \textsuperscript{\rm 4}Stanford University\\
    \textsuperscript{\rm 5}Massachusetts Institute of Technology \\
    leon@staufer.me, mickyang@seas.upenn.edu
}

\defcitealias{jiangMistral7B2023}{Mistral 7B}
\defcitealias{meinkeFrontierModelsAre}{Apollo}
\defcitealias{ustunAyaModelInstruction2024}{Cohere Aya}
\defcitealias{aisafetyinstituteAdvancedAIEvaluations2024}{UK AISI evals}
\defcitealias{teamGeminiFamilyHighly2024}{Gemini paper}
\defcitealias{grattafioriLlama3Herd2024}{Llama 3 paper}
\defcitealias{metrDetailsMETRsPreliminary2024}{METR o1}
\defcitealias{Claude3Model2024}{Claude 3 card}
\defcitealias{DetailsMETRsPreliminary2024}{METR Claude 3.5}
\defcitealias{teamGemma2Improving2024}{Gemma 2 paper}
\defcitealias{ukartificialintelligencesafetyinstitutePreDeploymentEvaluationAnthropics2024}{AISI Sonnet 3.5}
\defcitealias{openaiOpenAIO1System}{OpenAI o1 card}
\defcitealias{googleGemma2Model}{Gemma 2 card}
\defcitealias{geminiteamGemini15Unlocking2024}{Gemini 1.5 paper}
\defcitealias{ukartificialintelligencesafetyinstitutePreDeploymentEvaluationOpenAIs2024}{AISI o1}
\defcitealias{UpdateOurPreliminary2025}{METR update}
\defcitealias{deepseek-aiDeepSeekR1IncentivizingReasoning2025}{DeepSeek R1}
\defcitealias{openaiOpenAIO3miniSystem2025}{o3 mini card}
\defcitealias{deepseek-aiDeepSeekV3TechnicalReport2024}{DeepSeek V3}
\defcitealias{metrDetailsMETRsPreliminary2025}{METR o3 and o4-mini}
\defcitealias{anthropicSystemCardClaude2025}{Claude 4 card}
\defcitealias{geminiteamGemini25Pushing2025}{Gemini 2.5 paper}  
\defcitealias{InternationalJointTesting}{IJTE agent report}
\defcitealias{meta-llamaLlama3MODEL_CARDmd2024}{Llama 3 card}

\defcitealias{BrazilAIAct2023}{Brazil}
\defcitealias{chinaGenAI2025}{China}
\defcitealias{euAIAct}{EU}
\defcitealias{SGgov2024}{Singapore}
\defcitealias{SouthKoreaAIAct2024}{South Korea}
\defcitealias{USActionPlan2025}{USA}
\defcitealias{nyRAISE2025}{New York}
\defcitealias{bletchley2023}{Bletchley}
\defcitealias{intlaisafety2025}{Int'l Report}
\defcitealias{japanaisi2024}{Japan AISI}
\defcitealias{metr2024}{METR}
\defcitealias{nist2024}{NIST AI RMF}
\defcitealias{paris2025}{Paris}
\defcitealias{seoul2024}{Seoul}
\defcitealias{ukaisi2024}{UK AISI}
\defcitealias{anthropicResponsibleScalingPolicy2025}{Anthropic}
\defcitealias{FrontierSafetyFramework}{Google DeepMind}
\defcitealias{metaFrontierAIFramework}{Meta}
\defcitealias{microsoftFrontierGovernanceFramework2025}{Microsoft}
\defcitealias{openaiPreparednessFramework2025}{OpenAI}
\defcitealias{XAIRiskManagement}{xAI}

\begin{document}

\maketitle

\begin{abstract}
AI governance frameworks increasingly rely on audits, yet the results of their underlying evaluations require interpretation and context to be meaningfully informative. Even technically rigorous evaluations can offer little useful insight if reported selectively or obscurely. Current literature focuses primarily on technical best practices, but evaluations are an inherently sociotechnical process, and there is little guidance on reporting procedures and context. Through literature review, stakeholder interviews, and analysis of governance frameworks, we propose ``audit cards'' to make this context explicit. We identify six key types of contextual features to report and justify in audit cards: auditor identity, evaluation scope, methodology, resource access, process integrity, and review mechanisms. Through analysis of existing evaluation reports, we find significant variation in reporting practices, with most reports omitting crucial contextual information such as auditors' backgrounds, conflicts of interest, and the level and type of access to models. We also find that most existing regulations and frameworks lack guidance on rigorous reporting. In response to these shortcomings, we argue that audit cards can provide a structured format for reporting key claims alongside their justifications, enhancing transparency, facilitating proper interpretation, and establishing trust in reporting.
\end{abstract}

\section{Introduction} \label{sec:intro}

AI governance frameworks\footnote{This includes frontier AI safety policies from AI companies (e.g. OpenAI Preparedness Framework), regulations (e.g. US AI Action Plan, EU AI Act), and other norms (e.g. NIST AI Risk Management Framework). See \Cref{app:gov-selection} for a complete list.}
are being designed to increasingly rely on audits.\footnote{We understand audits as formalized evaluation processes of AI models. We refer to the specific evaluation procedure (e.g. benchmarking, red-teaming, uplift studies) underlying the audit as evaluations. See \Cref{app:eval-process} for an overview of key terms.}
Within these frameworks, audits are meant to (1) identify potential risks, (2) incentivize more responsive development practices, and (3) involve more stakeholders in the system deployment process.
However, audits can only fulfill this role effectively if the underlying evaluations are both technically rigorous and effectively integrated into decision-making \citep{hardy2024more}.
However, not all evaluations are equally rigorous (e.g., \citealp{raji2022outsider, birhaneAIAuditingBroken2024, mokander2023auditing, anderljungPubliclyAccountableFrontier2023, kolt2024responsible, casperBlackboxAccessInsufficient2024, reuelBetterBenchAssessingAI2024, li2024making}).

To date, much prior literature has focused on the technical side of audits (see \Cref{sec:related_work}). 
However, evaluations are never conducted in a vacuum \citep{wallach2025position}.
Their results are intimately shaped, not only by the technical methods employed, but by numerous non-technical details of their context and design (see \Cref{sec:eval-cards}). 
For example, while the research field has long understood the importance of conflict of interest disclosures for academic integrity (e.g., \citealp{knerr2020introduction}), auditor independence remains a consistent concern in the AI ecosystem \citep{costanza-chockWhoAuditsAuditors2022, evans2023who}.

Even technically rigorous evaluations can be uninformative or actively misleading if reported on selectively or obscurely \citep{ananny2018seeing}. 
This risk is magnified given the vested interest of AI developers in obtaining favorable audit results.
In AI and other fields, companies undergoing evaluations have strong incentives to game audits and influence reporting in potentially misleading ways (e.g., \citealp{krawiec2003cosmetic, lu2006does, marquis2016scrutiny}).

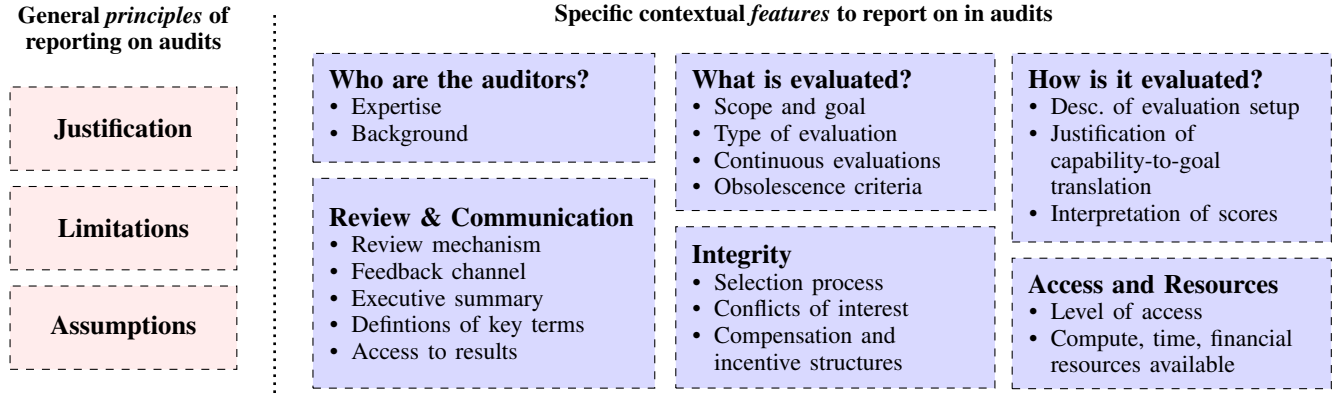
\begin{figure*}[tb]
    \centering
    \mbox{}\clap{
\begin{tikzpicture}[
    box/.style={draw, dashed, rectangle, minimum width=3cm, minimum height=1.1cm, align=center, fill=pink!30},
    bluebox/.style={draw, dashed, rectangle, text width=3.8cm, fill=blue!15, inner sep=6pt},
]

\def\xi{2.5}
\def\xii{7.3}
\def\xiii{11.75}
\def\yi{-0.5}

\node[align=center, font=\bfseries, scale=0.9] (principles) at (0,-0.2) {General \textit{principles} of\\reporting on audits};

\node[box] (justification) at (0,-1.5) {\textbf{Justification}};
\node[box, below=2mm of justification] (limitations)  {\textbf{Limitations}};
\node[box, below=2mm of limitations] (assumptions) {\textbf{Assumptions}};

\draw[dotted, line width=1pt] (2,0) -- (2,-5);

\node[align=center, font=\bfseries, scale=0.9] (features) at (9,0) {Specific contextual \textit{features} to report on in audits};

\node[bluebox, anchor=north west, text width=4.1cm] (evaluators) at (\xi,\yi) {
    \textbf{Who are the auditors?}
    \begin{itemize}[leftmargin=*, topsep=0pt, itemsep=0pt, parsep=0pt, before=\small]
        \item Expertise
        \item Background
    \end{itemize}
};
\node[bluebox, anchor=north west, below=2mm of evaluators, minimum height=2.77cm, text width=4.1cm] (review) {
    \textbf{Review \& Communication}
    \begin{itemize}[leftmargin=*, topsep=0pt, itemsep=0pt, parsep=0pt, before=\small]
        \item Review mechanism
        \item Feedback channel
        \item Executive summary
        \item Defintions of key terms
        \item Access to results
    \end{itemize}
};

\node[bluebox, anchor=north west] (target) at (\xii,\yi) {
    \textbf{What is evaluated?}
    \begin{itemize}[leftmargin=*, topsep=0pt, itemsep=0pt, parsep=0pt, before=\small]
        \item Scope and goal
        \item Type of evaluation
        \item Continuous evaluations
        \item Obsolescence criteria
    \end{itemize}
};

\node[bluebox, anchor=north west] (procedure) at (\xiii,\yi) {
    \textbf{How is it evaluated?}
    \begin{itemize}[leftmargin=*, topsep=0pt, itemsep=0pt, parsep=0pt, before=\small]
        \item Desc. of evaluation setup
        \item Justification of capability-to-goal translation
        \item Interpretation of scores
    \end{itemize}
};

\node[bluebox, anchor=north west, below=2mm of target, minimum height=2.13cm] (integrity) {
    \textbf{Integrity}
    \begin{itemize}[leftmargin=*, topsep=0pt, itemsep=0pt, parsep=0pt, before=\small]
        \item Selection process
        \item Conflicts of interest
        \item Compensation and incentive structures
    \end{itemize}
};

\node[bluebox, anchor=north west, below=2mm of procedure] (access) {
    \textbf{Access and Resources}
    \begin{itemize}[leftmargin=*, topsep=0pt, itemsep=0pt, parsep=0pt, before=\small]
        \item Level of access
        \item Compute, time, financial resources available
    \end{itemize}
};

\end{tikzpicture}
}
    \caption{Our audit card template requires reporting on three \textit{principles} and six \textit{features}. (Left) The three cross-cutting principles (justification, limitations, and assumptions) further transparency about the process behind an audit. (Right) The six features offer key methodological and contextual information. 
    See \Cref{sec:eval-cards} and \Cref{app:details-features} for details.}
    \label{fig:eval-card-features}
\end{figure*}

Thus, for AI audits to play a meaningful role in governance, evaluation reporting must be done in a way that minimizes the chance of omitting key contextual information. 
To address this challenge, we propose ``audit cards'': a structured reporting framework to document the context of an AI audit rather than the evaluated system itself (\Cref{fig:eval-card-features}).
Unlike model and system cards \citep{mitchellModelCardsModel2019}, which document system characteristics and capabilities, audit cards address the distinct transparency requirements of evaluation processes-specifically the contextual factors that shape how evaluations are conducted and reported.
They are designed to offer a standardized approach to reporting that enhances transparency and facilitates informed interpretations of results. This need not necessarily be reported in a separate audit card document, rather the relevant context can be added as sections in existing model and system cards. 

Overall, we make four key contributions: 
\begin{enumerate}[leftmargin=*]
    \item \textbf{Audit cards framework addressing evaluation transparency gaps:} Based on a survey of 28 prior works on AI evaluations (\Cref{sec:eval-cards}), we synthesise existing literature to propose a template and checklist (\Cref{app:audit_cards}) for AI audit cards that addresses transparency needs not covered by existing tools, including three cross-cutting principles (justification, limitations, and assumptions) and six key types of contextual information (auditor identity, evaluation scope, methodology, resource access, process integrity, and review mechanisms).
    \item \textbf{Survey of frontier AI evaluation reports:} We analyze the thoroughness of 24 existing evaluation reports for frontier systems (\Cref{sec:eval-reports}).
    \item \textbf{Survey of governance frameworks:} We analyze gaps between best reporting practices and 21 existing AI governance frameworks (\Cref{sec:governance}).
    \item \textbf{Stakeholder insights:} We interview and present perspectives from 10 expert stakeholders on AI evaluation reporting practices (\Cref{sec:interviews}).
\end{enumerate}

\section{Related work} \label{sec:related_work} %

\textbf{Transparency and reporting:}
In the past decade, frontier AI research and development have largely shifted from being predominantly academia-driven to predominantly industry-driven \citep{maslejArtificialIntelligenceIndex2024}. 
This shift has been accompanied by a decrease in transparency around proprietary state-of-the-art systems.
As such, increasing transparency and awareness has emerged as a key goal of AI governance (e.g., \citealp{felzmann2020towards, haresamudram2023three, winecoff2024improving, chan2024visibility, bommasani2024foundation, kolt2024responsible}).
One barrier to meaningful transparency is a lack of standards for reporting \citep{maslejArtificialIntelligenceIndex2024}. 
In response to this challenge, previous works have proposed data, model, system, agent, and usage ``cards'' \citep{pushkarna2022data, mitchellModelCardsModel2019, gursoy2022system, casper2025ai, wahle2023ai} for documenting key information within the AI ecosystem. 
In this paper, we build on past work by introducing the notion of an ``audit card,'' surveying what prior literature suggests they should contain, and analyzing the current state of reporting around audits. 

\textbf{AI audits:} Formal evaluations, also known as ``audits,'' of AI systems have been proposed as a key objective to facilitate transparency and scrutiny \citep{raji2022outsider, anderljungPubliclyAccountableFrontier2023, mokander2023auditing, li2024making, reuel2024open}. 
Meanwhile, audits are increasingly incorporated into frameworks for AI governance (e.g., \citealp{euAIAct}).
However, as we will show in \Cref{sec:governance}, current governance frameworks often lack substantial guidance for how to report on audits.

\textbf{Technically rigorous evaluations:}
AI systems are audited using a variety of approaches including case studies, benchmarks, red-teaming, and mechanistic analysis \citep{bengio2024international}. 
However, the science of evaluating AI systems is still nascent \citep{WeNeedScience}. 
Not all evaluations are equally technically rigorous, and their apparent outcomes can be highly sensitive to framing and design (e.g., \citealp{schaeffer2023emergent, burnellRethinkReportingEvaluation2023, khan2025randomness}).
Toward improved technical practices, \citet{reuelBetterBenchAssessingAI2024} outline a set of 46 best practices for rigorous AI benchmark design. 
Meanwhile, \citet{metr_task_standard_2024} work toward a portable standard for framing and conducting capability evaluations. 
However, unlike prior work, here we focus on rigor in reporting both key technical details and \textit{context} behind AI audits. 

\textbf{On the inherent sociotechnical nature of evaluations:} 
AI evaluations depend on technical tools to assess system properties and risks. 
However, they are never conducted in a vacuum; they are always embedded in a broader sociotechnical and political context. 
Meanwhile, the standards that systems are evaluated against are inherently based on subjective human values.
Thus, AI evaluations represent a social science measurement challenge \citep{wallach2025position}.
Aside from lacking technical soundness, evaluations can fail to be rigorous and serve the public's interest for a variety of nontechnical reasons. 
Prior works have emphasized the role of evaluation integrity and integration in ensuring meaningful oversight \citep{ojewale2024towards, raji2022outsider, sharkeyCausalFrameworkAI2024}.
By surveying prior literature on evaluation procedures (\Cref{sec:eval-cards}), analyzing current evaluation reports (\Cref{sec:eval-reports}), and analyzing evaluation frameworks (\Cref{sec:governance}), we make progress toward a more contextual and critical understanding of AI audits.

\section{What information do audit cards need for rigorous reporting?}\label{sec:eval-cards}

\textbf{Methodology:} We identify key components of audit cards through an initial literature review of work on AI audits and evaluations, transparency, technical evaluation design, and sociotechnical approaches.\footnote{We also refined our audit card template based on structured interviews with expert stakeholders. See \Cref{sec:interviews}.}
We selected 28 total papers and manually annotated them, scoring the extent to which the paper recommends that aspect of reporting: is it a \textit{major argument} of the paper (2), a \textit{minor argument} of the paper (1), or \textit{not mentioned} in the paper (0). See \Cref{app:annot-method} for further details on the methodology. 
In \Cref{tab:tab_literature_overview}, we summarize the perspectives from all 28 papers on the three principles and six features. It's aim is to be a comprehensive and exhaustive overview of audit features  based on the relevant literature.
Next, we expand on these principles and features. 
See \Cref{fig:eval-card-features} for an overview of an audit card and \Cref{app:eval-card-example} for an actionable audit card checklist that discusses the relative importance of principles and features for different stakeholders and auditing types.

\begin{table*}[th]
    \centering
    \setlength{\tabcolsep}{2mm} %
\small{
\begin{tabular}{l|rrr|rrrrrr}
\toprule
 & \multicolumn{3}{c}{Principles} & \multicolumn{6}{c}{Features} \\
Paper & Justif. & Assum. & Limit. & Who & What & How & Access & Integrity & Review \\
\midrule
\citealp{anderljungPubliclyAccountableFrontier2023} & \cellcolor{level1}1 & \cellcolor{level3}2 & \cellcolor{level3}2 & \cellcolor{level3}2 & \cellcolor{level3}2 & \cellcolor{level3}2 & \cellcolor{level3}2 & \cellcolor{level3}2 & \cellcolor{level3}2 \\
\citealp{WeNeedScience} & \cellcolor{level3}2 & \cellcolor{level3}2 & \cellcolor{level0}0 & \cellcolor{level0}0 & \cellcolor{level3}2 & \cellcolor{level3}2 & \cellcolor{level0}0 & \cellcolor{level0}0 & \cellcolor{level0}0 \\
\citealp{barnettDeclareJustifyExplicit2024} & \cellcolor{level3}2 & \cellcolor{level3}2 & \cellcolor{level0}0 & \cellcolor{level1}1 & \cellcolor{level0}0 & \cellcolor{level0}0 & \cellcolor{level0}0 & \cellcolor{level0}0 & \cellcolor{level0}0 \\
\citealp{barnettWhatAIEvaluations2024} & \cellcolor{level1}1 & \cellcolor{level3}2 & \cellcolor{level3}2 & \cellcolor{level0}0 & \cellcolor{level3}2 & \cellcolor{level3}2 & \cellcolor{level3}2 & \cellcolor{level3}2 & \cellcolor{level0}0 \\
\citealp{birhaneAIAuditingBroken2024} & \cellcolor{level3}2 & \cellcolor{level3}2 & \cellcolor{level3}2 & \cellcolor{level3}2 & \cellcolor{level3}2 & \cellcolor{level0}0 & \cellcolor{level1}1 & \cellcolor{level3}2 & \cellcolor{level3}2 \\
\citealp{bucknallStructuredAccessThirdParty} & \cellcolor{level0}0 & \cellcolor{level3}2 & \cellcolor{level3}2 & \cellcolor{level0}0 & \cellcolor{level3}2 & \cellcolor{level0}0 & \cellcolor{level3}2 & \cellcolor{level3}2 & \cellcolor{level3}2 \\
\citealp{burnellRethinkReportingEvaluation2023} & \cellcolor{level3}2 & \cellcolor{level3}2 & \cellcolor{level0}0 & \cellcolor{level0}0 & \cellcolor{level1}1 & \cellcolor{level3}2 & \cellcolor{level0}0 & \cellcolor{level0}0 & \cellcolor{level3}2 \\
\citealp{casperBlackboxAccessInsufficient2024} & \cellcolor{level1}1 & \cellcolor{level3}2 & \cellcolor{level0}0 & \cellcolor{level1}1 & \cellcolor{level0}0 & \cellcolor{level0}0 & \cellcolor{level3}2 & \cellcolor{level0}0 & \cellcolor{level1}1 \\
\citealp{changSurveyEvaluationLarge2023} & \cellcolor{level0}0 & \cellcolor{level3}2 & \cellcolor{level0}0 & \cellcolor{level3}2 & \cellcolor{level0}0 & \cellcolor{level3}2 & \cellcolor{level0}0 & \cellcolor{level0}0 & \cellcolor{level0}0 \\
\citealp{costanza-chockWhoAuditsAuditors2022} & \cellcolor{level3}2 & \cellcolor{level3}2 & \cellcolor{level0}0 & \cellcolor{level3}2 & \cellcolor{level1}1 & \cellcolor{level0}0 & \cellcolor{level3}2 & \cellcolor{level1}1 & \cellcolor{level3}2 \\
\citealp{dobbeHardChoicesArtificial2021} & \cellcolor{level3}2 & \cellcolor{level3}2 & \cellcolor{level3}2 & \cellcolor{level3}2 & \cellcolor{level1}1 & \cellcolor{level3}2 & \cellcolor{level0}0 & \cellcolor{level0}0 & \cellcolor{level3}2 \\
\citealp{dowDimensionsGenerativeAI2024} & \cellcolor{level0}0 & \cellcolor{level3}2 & \cellcolor{level0}0 & \cellcolor{level1}1 & \cellcolor{level3}2 & \cellcolor{level3}2 & \cellcolor{level0}0 & \cellcolor{level0}0 & \cellcolor{level0}0 \\
\citealp{erikssonCanWeTrust2025b} & \cellcolor{level3}2 & \cellcolor{level3}2 & \cellcolor{level1}1 & \cellcolor{level1}1 & \cellcolor{level3}2 & \cellcolor{level3}2 & \cellcolor{level0}0 & \cellcolor{level1}1 & \cellcolor{level0}0 \\
\citealp{gallifantTRIPODLLMReportingGuideline2025} & \cellcolor{level3}2 & \cellcolor{level3}2 & \cellcolor{level3}2 & \cellcolor{level3}2 & \cellcolor{level3}2 & \cellcolor{level3}2 & \cellcolor{level1}1 & \cellcolor{level3}2 & \cellcolor{level3}2 \\
\citealp{gebruDatasheetsDatasets2021} & \cellcolor{level1}1 & \cellcolor{level3}2 & \cellcolor{level3}2 & \cellcolor{level3}2 & \cellcolor{level1}1 & \cellcolor{level3}2 & \cellcolor{level3}2 & \cellcolor{level3}2 & \cellcolor{level3}2 \\
\citealp{hendrycksDevisingMLMetrics2024} & \cellcolor{level0}0 & \cellcolor{level0}0 & \cellcolor{level1}1 & \cellcolor{level0}0 & \cellcolor{level0}0 & \cellcolor{level3}2 & \cellcolor{level0}0 & \cellcolor{level0}0 & \cellcolor{level1}1 \\
\citealp{kolt2024responsible} & \cellcolor{level1}1 & \cellcolor{level1}1 & \cellcolor{level0}0 & \cellcolor{level3}2 & \cellcolor{level1}1 & \cellcolor{level0}0 & \cellcolor{level0}0 & \cellcolor{level3}2 & \cellcolor{level3}2 \\
\citealp{liangHolisticEvaluationLanguage2023} & \cellcolor{level1}1 & \cellcolor{level3}2 & \cellcolor{level0}0 & \cellcolor{level0}0 & \cellcolor{level3}2 & \cellcolor{level0}0 & \cellcolor{level0}0 & \cellcolor{level0}0 & \cellcolor{level0}0 \\
\citealp{metrTaskDevelopmentGuide} & \cellcolor{level0}0 & \cellcolor{level0}0 & \cellcolor{level0}0 & \cellcolor{level3}2 & \cellcolor{level3}2 & \cellcolor{level3}2 & \cellcolor{level0}0 & \cellcolor{level0}0 & \cellcolor{level3}2 \\
\citealp{mokanderAuditingLargeLanguage2024} & \cellcolor{level0}0 & \cellcolor{level3}2 & \cellcolor{level3}2 & \cellcolor{level1}1 & \cellcolor{level3}2 & \cellcolor{level1}1 & \cellcolor{level3}2 & \cellcolor{level3}2 & \cellcolor{level3}2 \\
\citealp{mukobiReasonsDoubtImpact2024} & \cellcolor{level0}0 & \cellcolor{level3}2 & \cellcolor{level0}0 & \cellcolor{level3}2 & \cellcolor{level3}2 & \cellcolor{level3}2 & \cellcolor{level3}2 & \cellcolor{level3}2 & \cellcolor{level1}1 \\
\citealp{ojewale2024towards} & \cellcolor{level1}1 & \cellcolor{level3}2 & \cellcolor{level0}0 & \cellcolor{level3}2 & \cellcolor{level1}1 & \cellcolor{level1}1 & \cellcolor{level3}2 & \cellcolor{level1}1 & \cellcolor{level1}1 \\
\citealp{rajiClosingAIAccountability2020} & \cellcolor{level0}0 & \cellcolor{level3}2 & \cellcolor{level3}2 & \cellcolor{level3}2 & \cellcolor{level3}2 & \cellcolor{level3}2 & \cellcolor{level1}1 & \cellcolor{level0}0 & \cellcolor{level1}1 \\
\citealp{reuelBetterBenchAssessingAI2024} & \cellcolor{level0}0 & \cellcolor{level3}2 & \cellcolor{level3}2 & \cellcolor{level3}2 & \cellcolor{level3}2 & \cellcolor{level3}2 & \cellcolor{level0}0 & \cellcolor{level0}0 & \cellcolor{level3}2 \\
\citealp{selbstFairnessAbstractionSociotechnical2019} & \cellcolor{level0}0 & \cellcolor{level3}2 & \cellcolor{level3}2 & \cellcolor{level3}2 & \cellcolor{level0}0 & \cellcolor{level3}2 & \cellcolor{level0}0 & \cellcolor{level0}0 & \cellcolor{level0}0 \\
\citealp{shevlaneModelEvaluationExtreme2023} & \cellcolor{level0}0 & \cellcolor{level3}2 & \cellcolor{level3}2 & \cellcolor{level3}2 & \cellcolor{level3}2 & \cellcolor{level3}2 & \cellcolor{level3}2 & \cellcolor{level3}2 & \cellcolor{level3}2 \\
\citealp{weidingerEvaluationScienceGenerative2025a} & \cellcolor{level3}2 & \cellcolor{level3}2 & \cellcolor{level3}2 & \cellcolor{level0}0 & \cellcolor{level3}2 & \cellcolor{level3}2 & \cellcolor{level0}0 & \cellcolor{level0}0 & \cellcolor{level1}1 \\
\citealp{zhanEvaluatologyScienceEngineering2024} & \cellcolor{level3}2 & \cellcolor{level3}2 & \cellcolor{level3}2 & \cellcolor{level1}1 & \cellcolor{level3}2 & \cellcolor{level3}2 & \cellcolor{level0}0 & \cellcolor{level0}0 & \cellcolor{level3}2 \\
\bottomrule
\end{tabular}
}

    \caption{\textbf{What contextual details do prior works say are key for audit reporting?} We divide reporting into three overarching principles (justifications, assumptions, limitations) and six key features (who, what, how, access, integrity, review). We score each paper's extent of recommendation of each these as 2 (major argument), 1 (minor argument), and 0 (no mention). Refer to \Cref{app:tab-lit-all-features} for our complete, more granular analysis of papers.}
    \label{tab:tab_literature_overview}
\end{table*}

\subsection{Overarching principles for transparent reporting}
The following three overarching principles behind auditing are key for transparency and methodological clarity. These apply across the entire auditing process and the six specific features.
We design audit cards to make it easier to adopt these principles throughout the auditing process. This, in return, enables accurate interpretation, facilitates constructive critique, and builds trust in the evaluation process.

\textbf{\textit{Justifications} as explicit arguments to support methodological choices, metric selection, and interpretative frameworks.} Explicit reporting on justifications enables process transparency. This justification can be through elements of a rigorous scientific process, such as explaining proxies and experimental coverage \citep{WeNeedScience}, or through standardized auditing procedures, such as detailed standards identification \citep{birhaneAIAuditingBroken2024, costanza-chockWhoAuditsAuditors2022}. 
    
\textbf{\textit{Assumptions} to articulate the premises underlying the evaluation design and analysis.}\footnote{This differs from justifications insofar that they answer ``Why did we choose this approach?" whereas assumptions answer ``What are we taking as given?"}
Reporting on assumptions clarifies the relationship (or lack thereof) between tests and real-world outcomes, presumptions about what constitutes ``good'' performance, and details about the threat models considered \citep{reuelBetterBenchAssessingAI2024, barnettDeclareJustifyExplicit2024, barnettWhatAIEvaluations2024, rajiClosingAIAccountability2020}.
    
\textbf{\textit{Limitations} to explicitly acknowledge the constraints that affect the validity, reliability, or generalizability of findings.} Reporting limitations clarifies constraints in data sampling, potential artifacts in the evaluation process, and boundaries of what auditors can reasonably claim based on the evaluation evidence \citep{mokanderAuditingLargeLanguage2024,barnettWhatAIEvaluations2024,reuelBetterBenchAssessingAI2024}.

\subsection{Specific features for contextual information} 
Aside from overarching principles, reporting on more specific contextual features of audits helps stakeholders assess trustworthiness. We compile recommendations from past literature into six features for reporting evaluation
context in audit cards.
For a more granular analysis, see \Cref{app:details-features}, where we further divide these six features into constituent aspects.

\begin{itemize}[leftmargin=*]

\item \textbf{Who are the auditors:} Reporting details of auditor identity facilitates trust and clarifies what potential biases might affect their work.
It includes reporting the \textit{expertise} (relevant domain knowledge, accreditations) \citep{anderljungPubliclyAccountableFrontier2023, metrTaskDevelopmentGuide, reuelBetterBenchAssessingAI2024} and \textit{background} (track record of conducting evaluations, positionality) of auditors \citep{costanza-chockWhoAuditsAuditors2022, gebruDatasheetsDatasets2021}. 
Given worries about the privacy of auditors, information about the background can be anonymized and aggregated.

\item \textbf{What is evaluated:} Reporting on evaluation goals enables clear interpretation of findings. This includes the \textit{scope} (e.g., model vs. scaffolding) and \textit{goal} (context of use) \citep{rajiClosingAIAccountability2020,birhaneAIAuditingBroken2024,shevlaneModelEvaluationExtreme2023}, \textit{type of evaluation} (e.g., capability vs. propensity), as the scope of the evaluation influences what the relevant level of access is \citep{metr_task_standard_2024, liangHolisticEvaluationLanguage2023}, as well as reporting on the evaluation \textit{obsolescence criteria} \citep{joaquinDeprecatingBenchmarksCriteria2025}.\footnote{Examples of obsolescence criteria could include substantial amounts of model fine-tuning, changes in system components, or changes in deployment context. This helps stakeholders understand the shelf life of the findings.} 
It also includes reporting whether the evaluation will be \textit{repeated}, and if so, at what time or after which conditions are met \citep{barnettWhatAIEvaluations2024, anderljungPubliclyAccountableFrontier2023}.

\item \textbf{How is it evaluated:} Explaining the underlying procedures for the evaluation, such as \textit{descriptions of the evaluation setup} with justifications of how a \textit{capability translates to evaluation goal} \citep{WeNeedScience,rajiClosingAIAccountability2020,reuelBetterBenchAssessingAI2024} and provide \textit{interpretation of scores} \citep{dobbeHardChoicesArtificial2021, selbstFairnessAbstractionSociotechnical2019}. The specific relevant aspects to report for this feature may change depending on the underlying evaluation. We therefore refer to the appropriate literature for the underlying evaluation for a complete list of relevant aspects to report and details on how to report on them, e.g. for benchmarks \citep{reuelBetterBenchAssessingAI2024}.

\item \textbf{Access and resources:} Report the constraints under which the evaluation was conducted to allow readers to critically assess the audit's limitations. This includes, what \textit{access} auditors have to a system (black/white box, with/without safeguards) and which \textit{resources available are available} (compute, time, funding) as this can have a large impact on the rigor of evaluations \citep{barnettWhatAIEvaluations2024, ojewale2024towards, casperBlackboxAccessInsufficient2024}.

\item \textbf{Integrity:} Accountability for the auditing process establishes trust in 
auditors and rules out potential conflicts of interest. This includes reporting transparently about the \textit{selection process} (how auditors were chosen) and \textit{compensation} (funding sources, incentive structures). These dynamics can significantly affect the evaluation process and are therefore crucial context \citep{birhaneAIAuditingBroken2024, costanza-chockWhoAuditsAuditors2022, gebruDatasheetsDatasets2021}.

\item \textbf{Review and communication:} Reports on if and what \textit{independent reviews} of audit methods facilitate accountability\footnote{We discuss issues of intellectual property and non-disclosure agreements (NDAs) standing in the way of internal review in \Cref{sec:interviews} and \Cref{sec:discussion}. This review can be performed by experts at the organization by cross-checking the final report, insofar as they were not involved in the evaluation beforehand.} \citep{reuelBetterBenchAssessingAI2024,anderljungPubliclyAccountableFrontier2023}. This further includes \textit{contact information} and \textit{avenues for corrections}, \textit{executive summaries} (covering the evaluation process and result for non-technical audiences), and \textit{clear definitions} of key terms \citep{gebruDatasheetsDatasets2021, bucknallStructuredAccessThirdParty, metr_task_standard_2024, reuelBetterBenchAssessingAI2024}.

\end{itemize}

Some of the above features may be considered confidential in certain contexts, with academic researchers potentially sharing different details than commercial entities \citep{mitchellModelCardsModel2019}. 
Recommendations about how and with whom audit cards should be shared are beyond the scope of this report.
However, redactions offer a simple, well-precedented approach for protecting sensitive information \citep{booneRedactingProprietaryInformation2015}.

\section{What do existing evaluation reports include?} \label{sec:eval-reports}

\textbf{Methodology:} To study the thoroughness of existing evaluation reports, we examined 24 reports produced between 2023 and 2025. 
We selected these 24 due to their focus on frontier systems, ensuring to include reports from both developers and third party organizations.   
We manually annotated each report using the 3 principles and 6 features from our audit card template (\Cref{sec:eval-cards}).
For each audit card component, we scored reports on a 0-2 scale for comprehensively reporting that component (2), minimally reporting that component (1), not reporting on this component at all (0). 
We assessed \textit{reporting thoroughness} only, not execution quality, making no judgments about how well evaluations were executed. 
This means, for example, transparently reporting that there was a lack of any review/feedback process and justifying this would receive a full score of 2. %
We summarize the findings in \Cref{tab:eval-reports}.

\begin{table*}[th]
    \centering
    \setlength{\tabcolsep}{2mm} %
\small{
\begin{tabular}{lr|rrr|rrrrrr|r}
\toprule
\multicolumn{2}{l}{Report \hspace{3em}Release date} & Justif. & Assum. & Limit. & Who & What & How & Access & Integrity & Review & Average \\
\midrule
Average & & 1.17 & 0.67 & 1.42 & 0.75 & 1.38 & 1.29 & 0.50 & 0.17 & 1.13 & 0.94 \\
\midrule
\citetalias{jiangMistral7B2023} & 10/23 & \cellcolor{level0}0 & \cellcolor{level0}0 & \cellcolor{level0}0 & \cellcolor{level0}0 & \cellcolor{level1}1 & \cellcolor{level1}1 & \cellcolor{level0}0 & \cellcolor{level0}0 & \cellcolor{level0}0 & 0.22 \\
\citetalias{meinkeFrontierModelsAre} & 01/24 & \cellcolor{level3}2 & \cellcolor{level1}1 & \cellcolor{level3}2 & \cellcolor{level0}0 & \cellcolor{level3}2 & \cellcolor{level3}2 & \cellcolor{level0}0 & \cellcolor{level0}0 & \cellcolor{level3}2 & 1.22 \\
\citetalias{ustunAyaModelInstruction2024} & 02/24 & \cellcolor{level3}2 & \cellcolor{level1}1 & \cellcolor{level3}2 & \cellcolor{level3}2 & \cellcolor{level3}2 & \cellcolor{level1}1 & \cellcolor{level0}0 & \cellcolor{level1}1 & \cellcolor{level1}1 & 1.33 \\
\citetalias{aisafetyinstituteAdvancedAIEvaluations2024} & 05/24 & \cellcolor{level1}1 & \cellcolor{level0}0 & \cellcolor{level1}1 & \cellcolor{level0}0 & \cellcolor{level1}1 & \cellcolor{level1}1 & \cellcolor{level0}0 & \cellcolor{level0}0 & \cellcolor{level3}2 & 0.67 \\
\citetalias{teamGeminiFamilyHighly2024} & 06/24 & \cellcolor{level1}1 & \cellcolor{level1}1 & \cellcolor{level1}1 & \cellcolor{level1}1 & \cellcolor{level1}1 & \cellcolor{level1}1 & \cellcolor{level1}1 & \cellcolor{level1}1 & \cellcolor{level1}1 & 1.00 \\
\citetalias{grattafioriLlama3Herd2024} & 07/24 & \cellcolor{level1}1 & \cellcolor{level1}1 & \cellcolor{level3}2 & \cellcolor{level3}2 & \cellcolor{level1}1 & \cellcolor{level3}2 & \cellcolor{level0}0 & \cellcolor{level0}0 & \cellcolor{level1}1 & 1.11 \\
\citetalias{meta-llamaLlama3MODEL_CARDmd2024} & 07/24 & \cellcolor{level1}1 & \cellcolor{level1}1 & \cellcolor{level1}1 & \cellcolor{level1}1 & \cellcolor{level1}1 & \cellcolor{level1}1 & \cellcolor{level0}0 & \cellcolor{level0}0 & \cellcolor{level1}1 & 0.78 \\
\citetalias{metrDetailsMETRsPreliminary2024} & 09/24 & \cellcolor{level1}1 & \cellcolor{level0}0 & \cellcolor{level1}1 & \cellcolor{level1}1 & \cellcolor{level1}1 & \cellcolor{level1}1 & \cellcolor{level1}1 & \cellcolor{level0}0 & \cellcolor{level1}1 & 0.78 \\
\citetalias{Claude3Model2024} & 10/24 & \cellcolor{level1}1 & \cellcolor{level1}1 & \cellcolor{level1}1 & \cellcolor{level1}1 & \cellcolor{level1}1 & \cellcolor{level1}1 & \cellcolor{level0}0 & \cellcolor{level0}0 & \cellcolor{level1}1 & 0.78 \\
\citetalias{DetailsMETRsPreliminary2024} & 10/24 & \cellcolor{level1}1 & \cellcolor{level0}0 & \cellcolor{level3}2 & \cellcolor{level1}1 & \cellcolor{level1}1 & \cellcolor{level1}1 & \cellcolor{level0}0 & \cellcolor{level0}0 & \cellcolor{level1}1 & 0.78 \\
\citetalias{teamGemma2Improving2024} & 10/24 & \cellcolor{level1}1 & \cellcolor{level0}0 & \cellcolor{level1}1 & \cellcolor{level0}0 & \cellcolor{level3}2 & \cellcolor{level1}1 & \cellcolor{level0}0 & \cellcolor{level0}0 & \cellcolor{level1}1 & 0.67 \\
\citetalias{ukartificialintelligencesafetyinstitutePreDeploymentEvaluationAnthropics2024} & 11/24 & \cellcolor{level3}2 & \cellcolor{level1}1 & \cellcolor{level3}2 & \cellcolor{level1}1 & \cellcolor{level3}2 & \cellcolor{level1}1 & \cellcolor{level1}1 & \cellcolor{level0}0 & \cellcolor{level1}1 & 1.22 \\
\citetalias{openaiOpenAIO1System} & 12/24 & \cellcolor{level1}1 & \cellcolor{level1}1 & \cellcolor{level1}1 & \cellcolor{level1}1 & \cellcolor{level1}1 & \cellcolor{level1}1 & \cellcolor{level1}1 & \cellcolor{level0}0 & \cellcolor{level1}1 & 0.89 \\
\citetalias{googleGemma2Model} & 12/24 & \cellcolor{level1}1 & \cellcolor{level0}0 & \cellcolor{level1}1 & \cellcolor{level0}0 & \cellcolor{level1}1 & \cellcolor{level1}1 & \cellcolor{level0}0 & \cellcolor{level0}0 & \cellcolor{level1}1 & 0.56 \\
\citetalias{geminiteamGemini15Unlocking2024} & 12/24 & \cellcolor{level1}1 & \cellcolor{level1}1 & \cellcolor{level3}2 & \cellcolor{level3}2 & \cellcolor{level1}1 & \cellcolor{level3}2 & \cellcolor{level1}1 & \cellcolor{level1}1 & \cellcolor{level3}2 & 1.44 \\
\citetalias{ukartificialintelligencesafetyinstitutePreDeploymentEvaluationOpenAIs2024} & 12/24 & \cellcolor{level3}2 & \cellcolor{level1}1 & \cellcolor{level3}2 & \cellcolor{level1}1 & \cellcolor{level3}2 & \cellcolor{level1}1 & \cellcolor{level1}1 & \cellcolor{level0}0 & \cellcolor{level1}1 & 1.22 \\
\citetalias{UpdateOurPreliminary2025} & 01/25 & \cellcolor{level1}1 & \cellcolor{level1}1 & \cellcolor{level3}2 & \cellcolor{level0}0 & \cellcolor{level1}1 & \cellcolor{level1}1 & \cellcolor{level0}0 & \cellcolor{level0}0 & \cellcolor{level3}2 & 0.89 \\
\citetalias{deepseek-aiDeepSeekR1IncentivizingReasoning2025} & 01/25 & \cellcolor{level0}0 & \cellcolor{level0}0 & \cellcolor{level0}0 & \cellcolor{level0}0 & \cellcolor{level1}1 & \cellcolor{level1}1 & \cellcolor{level0}0 & \cellcolor{level0}0 & \cellcolor{level0}0 & 0.22 \\
\citetalias{openaiOpenAIO3miniSystem2025} & 01/25 & \cellcolor{level1}1 & \cellcolor{level1}1 & \cellcolor{level1}1 & \cellcolor{level1}1 & \cellcolor{level1}1 & \cellcolor{level1}1 & \cellcolor{level1}1 & \cellcolor{level0}0 & \cellcolor{level1}1 & 0.89 \\
\citetalias{deepseek-aiDeepSeekV3TechnicalReport2024} & 02/25 & \cellcolor{level1}1 & \cellcolor{level0}0 & \cellcolor{level1}1 & \cellcolor{level0}0 & \cellcolor{level1}1 & \cellcolor{level1}1 & \cellcolor{level0}0 & \cellcolor{level0}0 & \cellcolor{level1}1 & 0.56 \\
\citetalias{metrDetailsMETRsPreliminary2025} & 04/25 & \cellcolor{level3}2 & \cellcolor{level1}1 & \cellcolor{level3}2 & \cellcolor{level1}1 & \cellcolor{level3}2 & \cellcolor{level3}2 & \cellcolor{level3}2 & \cellcolor{level0}0 & \cellcolor{level1}1 & 1.44 \\
\citetalias{anthropicSystemCardClaude2025} & 05/25 & \cellcolor{level3}2 & \cellcolor{level1}1 & \cellcolor{level3}2 & \cellcolor{level1}1 & \cellcolor{level3}2 & \cellcolor{level3}2 & \cellcolor{level1}1 & \cellcolor{level0}0 & \cellcolor{level1}1 & 1.33 \\
\citetalias{geminiteamGemini25Pushing2025} & 06/25 & \cellcolor{level1}1 & \cellcolor{level1}1 & \cellcolor{level3}2 & \cellcolor{level1}1 & \cellcolor{level3}2 & \cellcolor{level3}2 & \cellcolor{level1}1 & \cellcolor{level1}1 & \cellcolor{level3}2 & 1.44 \\
\citetalias{InternationalJointTesting} & 07/25 & \cellcolor{level1}1 & \cellcolor{level1}1 & \cellcolor{level3}2 & \cellcolor{level0}0 & \cellcolor{level3}2 & \cellcolor{level3}2 & \cellcolor{level1}1 & \cellcolor{level0}0 & \cellcolor{level1}1 & 1.11 \\
\bottomrule
\end{tabular}
}

\caption{\textbf{What contextual details do existing audit reports provide?} We score reports as providing comprehensive (2), minimal (1), or no information (0) for each component of an audit card. Table ordered by release date of the report.}
    \label{tab:eval-reports}
\end{table*}

\textbf{While all evaluation reports provide at least a general overview of procedures, the level of detail differs.} 
We find that audit reports consistently offer information on evaluation scope (average score 1.38) and procedures (1.29) used. 
However, contextual details are much less consistently reported (average 0.94, $\sigma=0.68$) both across organizations and sometimes within the same organization.
For example, Google DeepMind released four documents with significantly varying levels of contextual detail (average scores of 1.00, 0.67, 0.56, 1.44, and 1.44 in chronological order). %
Only some reports (e.g., \citet{meinkeFrontierModelsAre, teamGemma2Improving2024, ustunAyaModelInstruction2024}) offered detailed information on the scaffolding used to evaluate the system.

\textbf{Reports inconsistently address assumptions, limitations, and auditor information.} 
While justifications and limitations are more commonly acknowledged (22 of 24) than assumptions (16 of 24), they are typically discussed at a high level rather than relating to specific methodological choices. 
Ten reports discuss limitations minimally, while twelce do so comprehensively. 
However, there are notable qualitative differences even among more comprehensive discussions. 
For example, \citet{ustunAyaModelInstruction2024} discuss limitations in prompting techniques, language transferability, evaluation tool chaining, and reproducibility
in significant detail, whereas the OpenAI system card \citep{openaiOpenAIO1System} only consider limitations of \textit{what} was being evaluated (e.g. classified information or restricted data) rather than substantial limitations of the procedures. 
Regarding auditor information, 15 of 24 reports include auditor expertise details, but rarely their background (3 of 24), making an assessment of potential biases in the audit harder.

\textbf{Reports rarely disclose information about evaluation integrity, resources, review processes, and obsolescence criteria.} 
Only 4 of 24 reports detail organization and auditor selection, conflicts of interest, or contractual arrangements governing the evaluation process. Even in these cases, the report tended to be about the contractual details of human baselines that are part of the evaluation rather than the core auditing team itself. 
Only 11 of 24 reports specify which resources auditors had access to, including model access level, computational resources, and time constraints. 
Similarly, only eight reported on some form of review and ten on feedback mechanisms \citep{meinkeFrontierModelsAre,metrDetailsMETRsPreliminary2024,openaiOpenAIO1System,ukartificialintelligencesafetyinstitutePreDeploymentEvaluationAnthropics2024}. 
Seven report whether they underwent any form of peer review or quality assurance process before publication \citep[e.g.][]{teamGemini15Unlocking2024, ukartificialintelligencesafetyinstitutePreDeploymentEvaluationOpenAIs2024}.
Five specified an option for public feedback and corrections. 
Zero reports explicitly present obsolescence criteria.

\section{What guidance do current governance frameworks offer?} \label{sec:governance}

\begin{table*}[thb]
    \centering
    \setlength{\tabcolsep}{2mm} %
\small{
\begin{tabular}{l|rrr|rrrrrr}
\toprule
 & \multicolumn{3}{c}{Principles} & \multicolumn{6}{c}{Features} \\
Document & Justif. & Assum. & Limit. & Who & What & How & Access & Integrity  & Review \\
\midrule

 \multicolumn{10}{l}{\textbf{Regulations}} \\

\citetalias{BrazilAIAct2023} & 0 & 0 & 0 & 0 & 0 & 0 & 0 & 0 & 0 \\
\citetalias{chinaGenAI2025} & 0 & 0 & 0 & 0 & \cellcolor{level3}1 & \cellcolor{level3}1 & 0 & 0 & \cellcolor{level3}1 \\
\citetalias{euAIAct} & \cellcolor{level3}1 & \cellcolor{level3}1 & \cellcolor{level3}1 & \cellcolor{level3}1 & \cellcolor{level3}1 & \cellcolor{level3}1 & \cellcolor{level3}1 & \cellcolor{level3}1 & \cellcolor{level3}1 \\
\citetalias{SGgov2024} & 0 & 0 & 0 & \cellcolor{level3}1 & \cellcolor{level3}1 & \cellcolor{level3}1 & 0 & 0 & \cellcolor{level3}1 \\
\citetalias{SouthKoreaAIAct2024} & 0 & 0 & 0 & 0 & 0 & 0 & 0 & 0 & 0 \\
\citetalias{USActionPlan2025} & 0 & 0 & 0 & 0 & \cellcolor{level3}1 & 0 & 0 & 0 & 0 \\
\citetalias{nyRAISE2025} & 0 & 0 & 0 & 0 & 0 & \cellcolor{level3}1 & 0 & 0 & 0 \\

\midrule
\multicolumn{10}{l}{\textbf{Other norms}} \\
\citetalias{bletchley2023} & 0 & 0 & 0 & 0 & 0 & \cellcolor{level3}1 & 0 & 0 & \cellcolor{level3}1 \\
\citetalias{intlaisafety2025} & \cellcolor{level3}1 & \cellcolor{level3}1 & \cellcolor{level3}1 & \cellcolor{level3}1 & \cellcolor{level3}1 & \cellcolor{level3}1 & \cellcolor{level3}1 & \cellcolor{level3}1 & \cellcolor{level3}1 \\
\citetalias{japanaisi2024} & 0 & 0 & 0 & 0 & \cellcolor{level3}1 & 0 & 0 & 0 & 0 \\
\citetalias{metr2024} & \cellcolor{level3}1 & \cellcolor{level3}1 & \cellcolor{level3}1 & \cellcolor{level3}1 & \cellcolor{level3}1 & \cellcolor{level3}1 & 0 & \cellcolor{level3}1 & \cellcolor{level3}1 \\
\citetalias{nist2024} & \cellcolor{level3}1 & \cellcolor{level3}1 & \cellcolor{level3}1 & \cellcolor{level3}1 & \cellcolor{level3}1 & \cellcolor{level3}1 & 0 & \cellcolor{level3}1 & \cellcolor{level3}1 \\
\citetalias{paris2025} & 0 & 0 & 0 & 0 & 0 & 0 & 0 & 0 & 0 \\
\citetalias{seoul2024} & 0 & 0 & 0 & 0 & \cellcolor{level3}1 & 0 & 0 & 0 & 0 \\
\citetalias{ukaisi2024} & 0 & 0 & \cellcolor{level3}1 & \cellcolor{level3}1 & \cellcolor{level3}1 & \cellcolor{level3}1 & 0 & 0 & \cellcolor{level3}1 \\

\midrule
\multicolumn{10}{l}{\textbf{Company policies}} \\
\citetalias{anthropicResponsibleScalingPolicy2025} & \cellcolor{level3}1 & \cellcolor{level3}1 & 0 & 0 & \cellcolor{level3}1 & \cellcolor{level3}1 & 0 & \cellcolor{level3}1  & \cellcolor{level3}1 \\
\citetalias{FrontierSafetyFramework} & \cellcolor{level3}1 & 0 & \cellcolor{level3}1 & \cellcolor{level3}1 & \cellcolor{level3}1 & \cellcolor{level3}1 & 0 & \cellcolor{level3}1 & \cellcolor{level3}1 \\
\citetalias{metaFrontierAIFramework} & \cellcolor{level3}1 & 0 & \cellcolor{level3}1 & \cellcolor{level3}1 & \cellcolor{level3}1 & \cellcolor{level3}1 & 0 & \cellcolor{level3}1 & \cellcolor{level3}1 \\
\citetalias{microsoftFrontierGovernanceFramework2025} & \cellcolor{level3}1 & \cellcolor{level3}1 & \cellcolor{level3}1 & \cellcolor{level3}1 & \cellcolor{level3}1 & \cellcolor{level3}1 & 0 & 0 & \cellcolor{level3}1 \\
\citetalias{openaiPreparednessFramework2025} & \cellcolor{level3}1 & 0 & 0 & 0 & \cellcolor{level3}1 & \cellcolor{level3}1 & 0 & 0 & \cellcolor{level3}1 \\
\citetalias{XAIRiskManagement} & \cellcolor{level3}1 & 0 & \cellcolor{level3}1 & 0 & \cellcolor{level3}1 & \cellcolor{level3}1 & 0 & 0 & \cellcolor{level3}1 \\

\bottomrule

\end{tabular}
}

    \caption{\textbf{What guidance do existing governance frameworks provide for reporting?} We give binary scores of 1 (requires or recommends explicit reporting of this feature) or 0 (may mention issue's importance to the ecosystem but not specifically require or recommend its disclosure or reporting). For company policies we score disclosure of it. Citations and details for the specific documents are in \Cref{app:gov-selection}.}
    \label{tab:governance}
\end{table*}

\textbf{Methodology:} 
We analyzed key governance frameworks issued by institutions that are considered influential regarding technical evaluations for risks from state-of-the-art models. 
We selected these frameworks based on jurisdiction, institutional authority, and direct influence on model evaluations (see \Cref{app:gov-selection} for details). 
This included official regulations and policies (EU AI Act, US AI Action Plan), voluntary industry standards (NIST AI Risk Management Framework), and policies from major AI developers. 
We scored each framework on whether it either explicitly \textit{required} or \textit{recommended} reporting of each feature in our audit card template, as shown in \Cref{tab:governance}.

\textbf{Existing frameworks offer limited guidance on audit reporting.} 
We find existing frameworks consistently require that models are evaluated (treated synonymously with being assessed or tested) but are not at the level of detail to specifically require or even recommend evaluation reports state the features of the evaluation we have listed. This is especially true for these features: assumptions, limitations, process integrity matters, resources and access given to auditors involved.
This is somewhat expected, as governance frameworks for emerging technologies crystallize gradually \citep{linkov2018comparative}, and the supplementary lower-level details around evaluations in the form of soft law are still emerging.

\section{Challenges with auditing according to expert stakeholders} \label{sec:interviews}
 
\textbf{Methodology:} To more thoroughly understand audit reporting and perceived gaps between current and best practices, we interviewed 10 experts. 
We reached out to these experts based on their expertise and familiarity across the evaluation ecosystem.
The 10 experts come from a diverse range of private companies, non-profit organizations, and government bodies.
Interviewees include: evaluation designers (\textbf{P1}), evaluation developers (\textbf{P3, P4, P5}), evaluation report writers (\textbf{P1, P3, P5}), evaluation-related policy researchers (\textbf{P6, P7}), and those supporting evaluation delivery and standards development (\textbf{P8, P9}). 
Interview findings reveal insights that extend beyond that documented in existing literature.
To allow for candid discussions, we committed to working with the level of anonymity that was comfortable for experts. See \Cref{app:interviews} for methodological details, interviewees' backgrounds, interview questions and consent forms. 

\textbf{The quality of audits depends greatly on the auditor. Reporting on auditor details may become more critical as the evaluation market grows.} Currently, the quality and trustworthiness of evaluators are primarily assessed informally (\textbf{P1, P6, P7}) through impressions of auditors' work and reputation. Many evaluation reports currently lack detailed information about evaluator selection, training, and reporting methods for various reasons, including time constraints or the belief that technical details about the evaluation sufficiently indicate evaluation quality (\textbf{P3}). However, because third-party auditors' access to proprietary models is voluntary (\textbf{P1, P6}), regulatory gaps and market failures must be addressed to enable more reporting about auditor selection and engagement terms. Standardized evaluation reporting may become increasingly important as the market expands. Auditing regimes from other industries, such as finance, energy, and medicine, may offer guidance moving forward (\textbf{P9}; see also \citet{anderson-samwaysAIrelevantRegulatoryPrecedents2024}).

\textbf{Reporting about the evaluation timeline can help with industry's standardization efforts. The amount of time spent on designing and executing an evaluation can have a direct effect on the quality and thoroughness of the evaluation.} 
When audits begin and end, and the corresponding versions of model access at each checkpoint, can often be in tension with commercial pressures in companies to develop and deploy quickly (\textbf{P1, P4, P6}). Evaluation resources, including time, are often difficult to measure in a chaotic situation (\textbf{P3}). Standardization and guidance beforehand around auditing timeframes may be greatly appreciated by evaluators (\textbf{P1, P4, P5, P6}). In lieu of laws explicitly specifying timeframes for different stages of evaluation, reporting the time taken by auditors will be crucial to facilitate standardization and coordination. 

\textbf{Evaluation reports are currently used more by auditors than policymakers.} 
Currently, evaluation reports by both internal and external evaluators assume a primarily technical audience. Policymakers are constrained by time, technical understanding, regulatory state of affairs, and often read them at a high level (\textbf{P6, P9}). Thus, it seems that evaluation reports more directly inform the science of evaluations than policy, as they assist with reproducibility and elicitation. In the future, it will likely be useful for government authorities to check compliance with standards. 
Experts have also said that adapting features to a prioritization for each user group would be valuable (\textbf{P1, P5, P6, P8}). This may be the subject of future research.

\section{Discussion and recommendations} \label{sec:discussion}

\textbf{There are currently no established standards on how to report the context of audits.} 
The lack of standards leads to inconsistency in audit reporting practices.
Certain contextual features, such as reporting on the integrity of auditing processes, are particularly neglected. 
Lacking transparency around audit methodology and context impedes accountability in the AI ecosystem, where trust in evaluations is essential for informed decision-making by regulators, users, and the public.

\textbf{Audit cards address AI evaluation reporting gaps to aid methodological rigor and public transparency.}  
Current inconsistent reporting practices undermine evaluations' governance role, creating trust gaps between developers, auditors, regulators, and the public. By structuring the essential contextual information to be reported, audit cards enable meaningful interpretation of evaluation results. Countries vary in how they see evaluations—emphasizing scientific rigor or compliance verification, public transparency or government oversight. While this paper does not comment on which regulatory approach is superior, it does take the premise that evaluations are scientific exercises requiring public transparency for effective governance.

\textbf{Audit cards are one approach among several to increase transparency between developers, users, and other stakeholders.} 
Audit cards are designed to offer a structured checklist applicable to all types of assessment (see \Cref{app:eval-card-example}), but should be viewed as an instrument in a larger AI accountability toolkit.
The context they offer can be helpful to other mechanisms, including model and system cards, datasheets, and benchmark details \citep{mitchellModelCardsModel2019, gebruDatasheetsDatasets2021, reuelBetterBenchAssessingAI2024}. 
\\\\
\textbf{Limitations and concerns:} 
\begin{itemize}[leftmargin=*]
\item \textbf{Limited sample size and reporting details.} To enable consistent high-level analysis of the current ecosystem, we used simple 2 or 3 point scoring methods. However, this comes at the expense of granular details and nuance, differentiating between reports only to a limited degree.
This is particularly pronounced for our analysis of governance frameworks, where we use binary scores. 
Furthermore, our analysis, while extensive (28 academic papers, 20 evaluation reports, and 21 governance frameworks), is not exhaustive.
\item \textbf{Trade-offs between reporting quality and audit quality:} Implementing thorough reporting through audit cards requires additional time and effort from auditors who often already operate under significant resource constraints. The additional burden of comprehensive reporting may be challenging, particularly for smaller organizations or time-sensitive evaluations.
\item \textbf{Verification challenges:} Even with comprehensive reporting guidelines, there remain challenges in verifying the accuracy of disclosed information. However, fraud is not unique to AI. Other fields handle these challenges through formal scrutiny mechanisms and occasional legal action. %
Through explicit disclosure, audit cards can be a first step toward a more accountable governance regime. 
\end{itemize}

\vspace{2mm}
\textbf{Key near-term challenges with AI audit reporting include providing rigorous and standardized guidance.} 
In the absence of external, authoritative standards for reporting, stakeholders across the AI ecosystem will struggle to effectively compare and interpret audit results. 
The EU AI Act (including its Code of Practice) and various contributions from AISIs and NIST, international consortia, labs, and third-party evaluators have offered progress toward a shared understanding of rigor. 
However, future initiatives may benefit from a higher degree of specificity (e.g., templates) for reporting on the principles and features highlighted in this paper.

\textbf{Standardized reporting frameworks are key for trustworthy AI audits.} The path toward effective AI governance requires both technical innovation and procedural standardization. By establishing shared expectations for auditing transparency through structured reporting mechanisms like audit cards, policymakers can build a more accountable AI governance landscape. Such frameworks ensure evaluations are rigorous, not only in execution, but also in reporting. Ultimately, this enables stakeholders to make informed decisions based on transparent, comparable, and contextually rich information that serves the public interest.

\section*{Acknowledgments}
This research was supported by the ML Alignment \& Theory Scholars (MATS) Program, which provided funding for Leon Staufer and Mick Yang through research stipends. We also thank MATS and our research managers Juan Gil and Keivan Navaie for their organizational assistance and research support.

We express our gratitude to Michael Aird, Lily Stelling, as well as other members of the MATS cohort, participants in the FAR Labs discussion group, and others for their valuable feedback and suggestions throughout the development of this paper.

We are grateful to all participants who contributed through interviews. Their perspectives were instrumental in developing the frameworks presented.

\bibliography{main}

\clearpage

\appendix
\begin{appendices}
\section{Disambiguating auditing and the evaluations process} \label{app:eval-process}
\textbf{AI audits}, for the purpose of our analysis, refer to formalized evaluation processes of AI models or systems. By formalized, we mean that it is performed systematically and by an organization, i.e., not just being performed ad hoc by individuals. This kind of evaluation is a model audit, as opposed to a governance audit or an application audit \citep{mokanderAuditingLargeLanguage2024}. 

The \textbf{evaluation} is the underlying assessment run on the model during the audit. 
We account for the fact that evaluations can occur internally or externally, and pre- or post-deployment. Evaluations can assess capabilities, propensities, or risks. Evaluations can be in many forms, such as benchmarks, red-teaming, human uplift studies, and user studies. Our approach considers all these variants, as the contextual features we have identified are relevant to all. It is possible, that in some of these contexts certain features will be more important and should therefore be reported in more detail. We explore the importance of certain features further in \Cref{sec:discussion} and \Cref{app:details-features}. 

We conceptualize the evaluation process as comprising three main stages: (1) designing the underlying evaluation methodology to achieve its aims (this could be a benchmark, an evaluation suite, or a red-teaming exercise); (2) executing the evaluation and getting results; (3) publishing these results in some form. Our focus is to elaborate on stage (3), that is, what is to be included in these evaluation reports. 

When speaking of \textbf{evaluation reports}, we refer to internal or external documents including evaluation results (e.g., a model's results on a benchmark). Examples of evaluation reports include: \citep{openaiOpenAIO1System, meinkeFrontierModelsAre} as well as private evaluation reports produced by third-parties for AI companies and internal reports created by AI companies themselves.

\section{Literature review}\label{app:lit-review}
\subsection{Methodology Details}\label{app:annot-method}
We reviewed past work on ``AI auditing'', ``AI evaluations'', ``best practices'', ``science of evaluations'', ``evaluation context'' and ``reporting''. We then assessed each paper for relevance based off the titles. This only yielded eight papers. We thus expanded to adjacent fields such as transparency and AI auditing, technical evaluation design, and sociotechnical approaches. These fields integrate technical performance metrics with societal assessments and considerations of how technical performance interacts with real-world social, cultural, and institutional factors. 

We scored recommendations within each paper as \textit{major arguments} (2), \textit{minor arguments} (1), and \textit{no mention} (0). Scoring of (2) represents that the paper actively discussed and/or advocates for a feature, rather than mentioning the feature’s importance in passing (1). We chose our three-point scale to balance reliability and granularity. For all three annotations (papers, reports, and regulations), we deliberated on the appropriate scoring system to ensure it is objective, whilst not giving a false sense of detail.

Through four iterations, we refined which aspects were most salient, focusing particularly on aspects that were actionable for auditors. Our methodology included: (1) two annotators discussing discrepancies before establishing the scoring system, (2) iterating on granularity. We found fine-grained scoring unreliable at higher levels, opting instead for our three-tiered system, and (3) developing detailed scoring rubrics with concrete examples for each category. 

Finally, we categorized and prioritized these aspects based on input from experts. The mapping of aspects to features is shown in \Cref{app:mapping-features}. The final audit card is depicted in \Cref{fig:eval-card-features} and \Cref{app:audit_cards}.

The guidelines for annotation of audit reports in \Cref{sec:eval-reports} were as follows:
\begin{quote}
Focus is how well the features are \textit{reported} not how well they are argued/performed.

(0) no information / not mentioned. 

(1) some (even minimal) information. Example: ``based on conversations with global experts. [...] expert consultants" regarding expertise for biological evaluations in Claude 3 model card \citep[25]{Claude3Model2024}. 

(2) rich information, meaning you don't have any important open questions after reading. Example: “red team consists of experts in cybersecurity, [...], in addition to multilingual content specialists with backgrounds in integrity issues for specific geographic markets” regarding the expertise feature in Llama 3 research paper \citep[48]{grattafioriLlama3Herd2024}. 
\end{quote}

This granularity is commensurate with our aim: show from a bird's-eye view that current aspects of audit cards are inconsistently reported on, not provide a definitive assessment of individual audit reports. We find that awarding one point for a minimal amount of information is not an issue, as we want to be charitable in our assessment of reports while still illustrating a frequent lack of thorough reporting.

\subsection{List of papers}\label{app:paper-list}
\begin{enumerate}
    \item Towards Publicly Accountable Frontier LLMs \citep{anderljungPubliclyAccountableFrontier2023}
    \item Science of Evals \citep{WeNeedScience}
    \item Declare and Justify: Explicit assumptions in AI evaluations are necessary for effective regulation \citep{barnettDeclareJustifyExplicit2024}
    \item What AI evaluations for preventing catastrophic risks can and cannot do \citep{barnettWhatAIEvaluations2024}
    \item AI auditing: The Broken Bus on the Road to AI Accountability \citep{birhaneAIAuditingBroken2024}
    \item Structured Access for Third-Party Research \citep{bucknallStructuredAccessThirdParty}
    \item Rethink reporting of evaluation results in AI \citep{burnellRethinkReportingEvaluation2023}
    \item Black-Box Access is Insufficient for Rigorous AI Audits \citep{casperBlackboxAccessInsufficient2024}
    \item A Survey on Evaluation of Large Language Models \citep{changSurveyEvaluationLarge2023}
    \item Who Audits the Auditors? Recommendations from a field scan of the algorithmic auditing ecosystem \citep{costanza-chockWhoAuditsAuditors2022}
    \item Hard choices in artificial intelligence \citep{dobbeHardChoicesArtificial2021}
    \item Dimensions of Generative AI Evaluation Design \citep{dowDimensionsGenerativeAI2024}
    \item Can We Trust AI Benchmarks? \citep{erikssonCanWeTrust2025b}
    \item The TRIPOD-LLM Reporting Guideline for Studies Using Large Language Models \citep{gallifantTRIPODLLMReportingGuideline2025}
    \item Datasheets for Datasets \citep{gebruDatasheetsDatasets2021}
    \item Devising ML Metrics \citep{hendrycksDevisingMLMetrics2024}
    \item Responsible Reporting for Frontier {{AI}} Development \citep{kolt2024responsible}
    \item Holistic Evaluation of Language Models \citep{liangHolisticEvaluationLanguage2023}
    \item Task Development Guide \citep{metrTaskDevelopmentGuide}
    \item Auditing Large Language Models: A Three-Layered Approach \citep{mokanderAuditingLargeLanguage2024}
    \item Reasons to Doubt the Impact of AI Risk Evaluations \citep{mukobiReasonsDoubtImpact2024}
    \item Towards AI Accountability Infrastructure \citep{ojewale2024towards}
    \item Closing the AI Accountability Gap: Defining an End-to-End Framework for Internal Algorithmic Auditing \citep{rajiClosingAIAccountability2020}
    \item BetterBench \citep{reuelBetterBenchAssessingAI2024}
    \item Fairness and Abstraction in Sociotechnical Systems \citep{selbstFairnessAbstractionSociotechnical2019}
    \item Model evaluation for extreme risks \citep{shevlaneModelEvaluationExtreme2023}
    \item Toward an evaluation science for generative AI systems \citep{weidingerEvaluationScienceGenerative2025a}
    \item Evaluatology: The Science and Engineering of Evaluation \citep{zhanEvaluatologyScienceEngineering2024}
\end{enumerate}

\begin{table*}[ht]
    \centering
    \setlength{\tabcolsep}{1.2mm} %
\small{
\begin{tabular}{l|rrrrrrrrrrrrrrrrrrrrrrrr}
\toprule
Paper & 1 & 2 & 3 & 4 & 5 & 6 & 7 & 8 & 9 & 10 & 11 & 12 & 13 & 14 & 15 & 16 & 17 & 18 & 19 & 20 & 21 & 22 & 23 & 24 \\
\midrule
\citealp{anderljungPubliclyAccountableFrontier2023} & \cellcolor{level3}2 & \cellcolor{level0}0 & \cellcolor{level0}0 & \cellcolor{level3}2 & \cellcolor{level3}2 & \cellcolor{level3}2 & \cellcolor{level3}2 & \cellcolor{level3}2 & \cellcolor{level3}2 & \cellcolor{level3}2 & \cellcolor{level3}2 & \cellcolor{level3}2 & \cellcolor{level0}0 & \cellcolor{level3}2 & \cellcolor{level0}0 & \cellcolor{level0}0 & \cellcolor{level3}2 & \cellcolor{level3}2 & \cellcolor{level0}0 & \cellcolor{level0}0 & \cellcolor{level3}2 & \cellcolor{level1}1 & \cellcolor{level3}2 & \cellcolor{level3}2 \\
\citealp{WeNeedScience} & \cellcolor{level0}0 & \cellcolor{level0}0 & \cellcolor{level0}0 & \cellcolor{level0}0 & \cellcolor{level0}0 & \cellcolor{level3}2 & \cellcolor{level0}0 & \cellcolor{level0}0 & \cellcolor{level0}0 & \cellcolor{level0}0 & \cellcolor{level3}2 & \cellcolor{level0}0 & \cellcolor{level0}0 & \cellcolor{level0}0 & \cellcolor{level0}0 & \cellcolor{level0}0 & \cellcolor{level0}0 & \cellcolor{level0}0 & \cellcolor{level0}0 & \cellcolor{level0}0 & \cellcolor{level3}2 & \cellcolor{level3}2 & \cellcolor{level0}0 & \cellcolor{level0}0 \\
\citealp{barnettDeclareJustifyExplicit2024} & \cellcolor{level0}0 & \cellcolor{level0}0 & \cellcolor{level0}0 & \cellcolor{level0}0 & \cellcolor{level1}1 & \cellcolor{level0}0 & \cellcolor{level0}0 & \cellcolor{level0}0 & \cellcolor{level0}0 & \cellcolor{level0}0 & \cellcolor{level0}0 & \cellcolor{level1}1 & \cellcolor{level0}0 & \cellcolor{level0}0 & \cellcolor{level3}2 & \cellcolor{level0}0 & \cellcolor{level0}0 & \cellcolor{level0}0 & \cellcolor{level0}0 & \cellcolor{level0}0 & \cellcolor{level1}1 & \cellcolor{level3}2 & \cellcolor{level0}0 & \cellcolor{level0}0 \\
\citealp{barnettWhatAIEvaluations2024} & \cellcolor{level3}2 & \cellcolor{level3}2 & \cellcolor{level0}0 & \cellcolor{level1}1 & \cellcolor{level0}0 & \cellcolor{level0}0 & \cellcolor{level0}0 & \cellcolor{level0}0 & \cellcolor{level0}0 & \cellcolor{level0}0 & \cellcolor{level3}2 & \cellcolor{level0}0 & \cellcolor{level0}0 & \cellcolor{level0}0 & \cellcolor{level0}0 & \cellcolor{level3}2 & \cellcolor{level3}2 & \cellcolor{level0}0 & \cellcolor{level0}0 & \cellcolor{level0}0 & \cellcolor{level3}2 & \cellcolor{level1}1 & \cellcolor{level3}2 & \cellcolor{level0}0 \\
\citealp{birhaneAIAuditingBroken2024} & \cellcolor{level1}1 & \cellcolor{level3}2 & \cellcolor{level0}0 & \cellcolor{level0}0 & \cellcolor{level1}1 & \cellcolor{level3}2 & \cellcolor{level0}0 & \cellcolor{level0}0 & \cellcolor{level0}0 & \cellcolor{level0}0 & \cellcolor{level0}0 & \cellcolor{level3}2 & \cellcolor{level0}0 & \cellcolor{level0}0 & \cellcolor{level3}2 & \cellcolor{level0}0 & \cellcolor{level3}2 & \cellcolor{level3}2 & \cellcolor{level0}0 & \cellcolor{level0}0 & \cellcolor{level3}2 & \cellcolor{level3}2 & \cellcolor{level3}2 & \cellcolor{level3}2 \\
\citealp{bucknallStructuredAccessThirdParty} & \cellcolor{level3}2 & \cellcolor{level3}2 & \cellcolor{level0}0 & \cellcolor{level3}2 & \cellcolor{level0}0 & \cellcolor{level3}2 & \cellcolor{level3}2 & \cellcolor{level0}0 & \cellcolor{level0}0 & \cellcolor{level0}0 & \cellcolor{level0}0 & \cellcolor{level0}0 & \cellcolor{level3}2 & \cellcolor{level0}0 & \cellcolor{level0}0 & \cellcolor{level0}0 & \cellcolor{level3}2 & \cellcolor{level0}0 & \cellcolor{level0}0 & \cellcolor{level0}0 & \cellcolor{level3}2 & \cellcolor{level0}0 & \cellcolor{level3}2 & \cellcolor{level3}2 \\
\citealp{burnellRethinkReportingEvaluation2023} & \cellcolor{level0}0 & \cellcolor{level0}0 & \cellcolor{level0}0 & \cellcolor{level0}0 & \cellcolor{level0}0 & \cellcolor{level1}1 & \cellcolor{level3}2 & \cellcolor{level0}0 & \cellcolor{level0}0 & \cellcolor{level1}1 & \cellcolor{level3}2 & \cellcolor{level0}0 & \cellcolor{level0}0 & \cellcolor{level3}2 & \cellcolor{level3}2 & \cellcolor{level3}2 & \cellcolor{level0}0 & \cellcolor{level0}0 & \cellcolor{level1}1 & \cellcolor{level0}0 & \cellcolor{level3}2 & \cellcolor{level3}2 & \cellcolor{level0}0 & \cellcolor{level1}1 \\
\citealp{casperBlackboxAccessInsufficient2024} & \cellcolor{level3}2 & \cellcolor{level0}0 & \cellcolor{level1}1 & \cellcolor{level0}0 & \cellcolor{level0}0 & \cellcolor{level0}0 & \cellcolor{level0}0 & \cellcolor{level0}0 & \cellcolor{level0}0 & \cellcolor{level0}0 & \cellcolor{level0}0 & \cellcolor{level0}0 & \cellcolor{level0}0 & \cellcolor{level0}0 & \cellcolor{level0}0 & \cellcolor{level0}0 & \cellcolor{level0}0 & \cellcolor{level0}0 & \cellcolor{level0}0 & \cellcolor{level0}0 & \cellcolor{level3}2 & \cellcolor{level1}1 & \cellcolor{level0}0 & \cellcolor{level1}1 \\
\citealp{changSurveyEvaluationLarge2023} & \cellcolor{level0}0 & \cellcolor{level0}0 & \cellcolor{level0}0 & \cellcolor{level0}0 & \cellcolor{level1}1 & \cellcolor{level0}0 & \cellcolor{level0}0 & \cellcolor{level0}0 & \cellcolor{level3}2 & \cellcolor{level0}0 & \cellcolor{level0}0 & \cellcolor{level3}2 & \cellcolor{level0}0 & \cellcolor{level0}0 & \cellcolor{level3}2 & \cellcolor{level0}0 & \cellcolor{level0}0 & \cellcolor{level0}0 & \cellcolor{level0}0 & \cellcolor{level0}0 & \cellcolor{level1}1 & \cellcolor{level0}0 & \cellcolor{level0}0 & \cellcolor{level0}0 \\
\citealp{costanza-chockWhoAuditsAuditors2022} & \cellcolor{level3}2 & \cellcolor{level1}1 & \cellcolor{level0}0 & \cellcolor{level0}0 & \cellcolor{level3}2 & \cellcolor{level1}1 & \cellcolor{level0}0 & \cellcolor{level1}1 & \cellcolor{level0}0 & \cellcolor{level0}0 & \cellcolor{level0}0 & \cellcolor{level1}1 & \cellcolor{level0}0 & \cellcolor{level0}0 & \cellcolor{level0}0 & \cellcolor{level0}0 & \cellcolor{level0}0 & \cellcolor{level3}2 & \cellcolor{level0}0 & \cellcolor{level0}0 & \cellcolor{level3}2 & \cellcolor{level3}2 & \cellcolor{level0}0 & \cellcolor{level3}2 \\
\citealp{dobbeHardChoicesArtificial2021} & \cellcolor{level0}0 & \cellcolor{level0}0 & \cellcolor{level0}0 & \cellcolor{level0}0 & \cellcolor{level0}0 & \cellcolor{level0}0 & \cellcolor{level0}0 & \cellcolor{level0}0 & \cellcolor{level0}0 & \cellcolor{level0}0 & \cellcolor{level3}2 & \cellcolor{level0}0 & \cellcolor{level0}0 & \cellcolor{level3}2 & \cellcolor{level3}2 & \cellcolor{level0}0 & \cellcolor{level3}2 & \cellcolor{level3}2 & \cellcolor{level1}1 & \cellcolor{level3}2 & \cellcolor{level3}2 & \cellcolor{level3}2 & \cellcolor{level1}1 & \cellcolor{level3}2 \\
\citealp{dowDimensionsGenerativeAI2024} & \cellcolor{level0}0 & \cellcolor{level0}0 & \cellcolor{level1}1 & \cellcolor{level0}0 & \cellcolor{level1}1 & \cellcolor{level3}2 & \cellcolor{level0}0 & \cellcolor{level0}0 & \cellcolor{level3}2 & \cellcolor{level0}0 & \cellcolor{level1}1 & \cellcolor{level1}1 & \cellcolor{level0}0 & \cellcolor{level0}0 & \cellcolor{level0}0 & \cellcolor{level0}0 & \cellcolor{level0}0 & \cellcolor{level1}1 & \cellcolor{level0}0 & \cellcolor{level0}0 & \cellcolor{level3}2 & \cellcolor{level0}0 & \cellcolor{level3}2 & \cellcolor{level0}0 \\
\citealp{erikssonCanWeTrust2025b} & \cellcolor{level0}0 & \cellcolor{level1}1 & \cellcolor{level0}0 & \cellcolor{level0}0 & \cellcolor{level1}1 & \cellcolor{level3}2 & \cellcolor{level0}0 & \cellcolor{level0}0 & \cellcolor{level1}1 & \cellcolor{level0}0 & \cellcolor{level3}2 & \cellcolor{level1}1 & \cellcolor{level0}0 & \cellcolor{level1}1 & \cellcolor{level3}2 & \cellcolor{level3}2 & \cellcolor{level1}1 & \cellcolor{level0}0 & \cellcolor{level3}2 & \cellcolor{level3}2 & \cellcolor{level3}2 & \cellcolor{level3}2 & \cellcolor{level0}0 & \cellcolor{level0}0 \\
\citealp{gallifantTRIPODLLMReportingGuideline2025} & \cellcolor{level0}0 & \cellcolor{level0}0 & \cellcolor{level0}0 & \cellcolor{level1}1 & \cellcolor{level3}2 & \cellcolor{level3}2 & \cellcolor{level3}2 & \cellcolor{level3}2 & \cellcolor{level1}1 & \cellcolor{level0}0 & \cellcolor{level3}2 & \cellcolor{level1}1 & \cellcolor{level0}0 & \cellcolor{level3}2 & \cellcolor{level1}1 & \cellcolor{level1}1 & \cellcolor{level3}2 & \cellcolor{level0}0 & \cellcolor{level1}1 & \cellcolor{level0}0 & \cellcolor{level3}2 & \cellcolor{level3}2 & \cellcolor{level3}2 & \cellcolor{level3}2 \\
\citealp{gebruDatasheetsDatasets2021} & \cellcolor{level3}2 & \cellcolor{level1}1 & \cellcolor{level1}1 & \cellcolor{level0}0 & \cellcolor{level1}1 & \cellcolor{level1}1 & \cellcolor{level3}2 & \cellcolor{level3}2 & \cellcolor{level3}2 & \cellcolor{level1}1 & \cellcolor{level0}0 & \cellcolor{level1}1 & \cellcolor{level3}2 & \cellcolor{level3}2 & \cellcolor{level3}2 & \cellcolor{level3}2 & \cellcolor{level3}2 & \cellcolor{level3}2 & \cellcolor{level0}0 & \cellcolor{level0}0 & \cellcolor{level1}1 & \cellcolor{level1}1 & \cellcolor{level0}0 & \cellcolor{level3}2 \\
\citealp{hendrycksDevisingMLMetrics2024} & \cellcolor{level0}0 & \cellcolor{level0}0 & \cellcolor{level0}0 & \cellcolor{level0}0 & \cellcolor{level0}0 & \cellcolor{level0}0 & \cellcolor{level0}0 & \cellcolor{level0}0 & \cellcolor{level1}1 & \cellcolor{level1}1 & \cellcolor{level3}2 & \cellcolor{level0}0 & \cellcolor{level0}0 & \cellcolor{level0}0 & \cellcolor{level0}0 & \cellcolor{level0}0 & \cellcolor{level1}1 & \cellcolor{level0}0 & \cellcolor{level0}0 & \cellcolor{level0}0 & \cellcolor{level0}0 & \cellcolor{level0}0 & \cellcolor{level0}0 & \cellcolor{level1}1 \\
\citealp{kolt2024responsible} & \cellcolor{level0}0 & \cellcolor{level3}2 & \cellcolor{level0}0 & \cellcolor{level0}0 & \cellcolor{level3}2 & \cellcolor{level0}0 & \cellcolor{level3}2 & \cellcolor{level0}0 & \cellcolor{level0}0 & \cellcolor{level0}0 & \cellcolor{level0}0 & \cellcolor{level0}0 & \cellcolor{level0}0 & \cellcolor{level0}0 & \cellcolor{level0}0 & \cellcolor{level0}0 & \cellcolor{level0}0 & \cellcolor{level0}0 & \cellcolor{level1}1 & \cellcolor{level0}0 & \cellcolor{level1}1 & \cellcolor{level1}1 & \cellcolor{level1}1 & \cellcolor{level3}2 \\
\citealp{liangHolisticEvaluationLanguage2023} & \cellcolor{level0}0 & \cellcolor{level0}0 & \cellcolor{level0}0 & \cellcolor{level0}0 & \cellcolor{level0}0 & \cellcolor{level0}0 & \cellcolor{level0}0 & \cellcolor{level0}0 & \cellcolor{level0}0 & \cellcolor{level0}0 & \cellcolor{level0}0 & \cellcolor{level0}0 & \cellcolor{level0}0 & \cellcolor{level0}0 & \cellcolor{level0}0 & \cellcolor{level0}0 & \cellcolor{level0}0 & \cellcolor{level0}0 & \cellcolor{level3}2 & \cellcolor{level0}0 & \cellcolor{level3}2 & \cellcolor{level1}1 & \cellcolor{level0}0 & \cellcolor{level0}0 \\
\citealp{metrTaskDevelopmentGuide} & \cellcolor{level0}0 & \cellcolor{level0}0 & \cellcolor{level0}0 & \cellcolor{level0}0 & \cellcolor{level3}2 & \cellcolor{level3}2 & \cellcolor{level3}2 & \cellcolor{level0}0 & \cellcolor{level1}1 & \cellcolor{level3}2 & \cellcolor{level3}2 & \cellcolor{level3}2 & \cellcolor{level3}2 & \cellcolor{level0}0 & \cellcolor{level0}0 & \cellcolor{level0}0 & \cellcolor{level0}0 & \cellcolor{level0}0 & \cellcolor{level0}0 & \cellcolor{level0}0 & \cellcolor{level0}0 & \cellcolor{level0}0 & \cellcolor{level0}0 & \cellcolor{level0}0 \\
\citealp{mokanderAuditingLargeLanguage2024} & \cellcolor{level3}2 & \cellcolor{level3}2 & \cellcolor{level0}0 & \cellcolor{level0}0 & \cellcolor{level1}1 & \cellcolor{level3}2 & \cellcolor{level3}2 & \cellcolor{level0}0 & \cellcolor{level0}0 & \cellcolor{level0}0 & \cellcolor{level1}1 & \cellcolor{level1}1 & \cellcolor{level0}0 & \cellcolor{level0}0 & \cellcolor{level0}0 & \cellcolor{level0}0 & \cellcolor{level3}2 & \cellcolor{level1}1 & \cellcolor{level3}2 & \cellcolor{level0}0 & \cellcolor{level3}2 & \cellcolor{level0}0 & \cellcolor{level3}2 & \cellcolor{level3}2 \\
\citealp{mukobiReasonsDoubtImpact2024} & \cellcolor{level3}2 & \cellcolor{level3}2 & \cellcolor{level3}2 & \cellcolor{level3}2 & \cellcolor{level0}0 & \cellcolor{level3}2 & \cellcolor{level0}0 & \cellcolor{level3}2 & \cellcolor{level0}0 & \cellcolor{level0}0 & \cellcolor{level3}2 & \cellcolor{level0}0 & \cellcolor{level0}0 & \cellcolor{level1}1 & \cellcolor{level0}0 & \cellcolor{level0}0 & \cellcolor{level0}0 & \cellcolor{level0}0 & \cellcolor{level1}1 & \cellcolor{level0}0 & \cellcolor{level3}2 & \cellcolor{level0}0 & \cellcolor{level3}2 & \cellcolor{level1}1 \\
\citealp{ojewale2024towards} & \cellcolor{level3}2 & \cellcolor{level1}1 & \cellcolor{level1}1 & \cellcolor{level0}0 & \cellcolor{level1}1 & \cellcolor{level1}1 & \cellcolor{level1}1 & \cellcolor{level0}0 & \cellcolor{level1}1 & \cellcolor{level1}1 & \cellcolor{level1}1 & \cellcolor{level1}1 & \cellcolor{level0}0 & \cellcolor{level0}0 & \cellcolor{level0}0 & \cellcolor{level0}0 & \cellcolor{level0}0 & \cellcolor{level3}2 & \cellcolor{level1}1 & \cellcolor{level0}0 & \cellcolor{level3}2 & \cellcolor{level1}1 & \cellcolor{level0}0 & \cellcolor{level1}1 \\
\citealp{rajiClosingAIAccountability2020} & \cellcolor{level1}1 & \cellcolor{level0}0 & \cellcolor{level0}0 & \cellcolor{level0}0 & \cellcolor{level0}0 & \cellcolor{level3}2 & \cellcolor{level0}0 & \cellcolor{level0}0 & \cellcolor{level3}2 & \cellcolor{level0}0 & \cellcolor{level3}2 & \cellcolor{level1}1 & \cellcolor{level0}0 & \cellcolor{level0}0 & \cellcolor{level3}2 & \cellcolor{level0}0 & \cellcolor{level3}2 & \cellcolor{level3}2 & \cellcolor{level3}2 & \cellcolor{level3}2 & \cellcolor{level3}2 & \cellcolor{level0}0 & \cellcolor{level0}0 & \cellcolor{level1}1 \\
\citealp{reuelBetterBenchAssessingAI2024} & \cellcolor{level0}0 & \cellcolor{level0}0 & \cellcolor{level0}0 & \cellcolor{level0}0 & \cellcolor{level3}2 & \cellcolor{level0}0 & \cellcolor{level3}2 & \cellcolor{level0}0 & \cellcolor{level0}0 & \cellcolor{level3}2 & \cellcolor{level3}2 & \cellcolor{level3}2 & \cellcolor{level3}2 & \cellcolor{level3}2 & \cellcolor{level3}2 & \cellcolor{level0}0 & \cellcolor{level3}2 & \cellcolor{level0}0 & \cellcolor{level0}0 & \cellcolor{level0}0 & \cellcolor{level1}1 & \cellcolor{level0}0 & \cellcolor{level3}2 & \cellcolor{level3}2 \\
\citealp{selbstFairnessAbstractionSociotechnical2019} & \cellcolor{level0}0 & \cellcolor{level0}0 & \cellcolor{level0}0 & \cellcolor{level0}0 & \cellcolor{level0}0 & \cellcolor{level0}0 & \cellcolor{level0}0 & \cellcolor{level0}0 & \cellcolor{level3}2 & \cellcolor{level0}0 & \cellcolor{level3}2 & \cellcolor{level1}1 & \cellcolor{level0}0 & \cellcolor{level3}2 & \cellcolor{level3}2 & \cellcolor{level0}0 & \cellcolor{level3}2 & \cellcolor{level3}2 & \cellcolor{level0}0 & \cellcolor{level3}2 & \cellcolor{level3}2 & \cellcolor{level0}0 & \cellcolor{level0}0 & \cellcolor{level0}0 \\
\citealp{shevlaneModelEvaluationExtreme2023} & \cellcolor{level3}2 & \cellcolor{level3}2 & \cellcolor{level0}0 & \cellcolor{level0}0 & \cellcolor{level3}2 & \cellcolor{level3}2 & \cellcolor{level0}0 & \cellcolor{level0}0 & \cellcolor{level0}0 & \cellcolor{level0}0 & \cellcolor{level3}2 & \cellcolor{level0}0 & \cellcolor{level0}0 & \cellcolor{level0}0 & \cellcolor{level0}0 & \cellcolor{level1}1 & \cellcolor{level3}2 & \cellcolor{level0}0 & \cellcolor{level0}0 & \cellcolor{level0}0 & \cellcolor{level3}2 & \cellcolor{level0}0 & \cellcolor{level3}2 & \cellcolor{level3}2 \\
\citealp{weidingerEvaluationScienceGenerative2025a} & \cellcolor{level0}0 & \cellcolor{level0}0 & \cellcolor{level0}0 & \cellcolor{level0}0 & \cellcolor{level0}0 & \cellcolor{level3}2 & \cellcolor{level1}1 & \cellcolor{level0}0 & \cellcolor{level0}0 & \cellcolor{level0}0 & \cellcolor{level3}2 & \cellcolor{level0}0 & \cellcolor{level0}0 & \cellcolor{level3}2 & \cellcolor{level0}0 & \cellcolor{level0}0 & \cellcolor{level3}2 & \cellcolor{level0}0 & \cellcolor{level3}2 & \cellcolor{level1}1 & \cellcolor{level3}2 & \cellcolor{level3}2 & \cellcolor{level1}1 & \cellcolor{level0}0 \\
\citealp{zhanEvaluatologyScienceEngineering2024} & \cellcolor{level0}0 & \cellcolor{level0}0 & \cellcolor{level0}0 & \cellcolor{level0}0 & \cellcolor{level1}1 & \cellcolor{level3}2 & \cellcolor{level3}2 & \cellcolor{level0}0 & \cellcolor{level0}0 & \cellcolor{level0}0 & \cellcolor{level3}2 & \cellcolor{level0}0 & \cellcolor{level0}0 & \cellcolor{level3}2 & \cellcolor{level3}2 & \cellcolor{level3}2 & \cellcolor{level3}2 & \cellcolor{level0}0 & \cellcolor{level0}0 & \cellcolor{level0}0 & \cellcolor{level3}2 & \cellcolor{level3}2 & \cellcolor{level0}0 & \cellcolor{level0}0 \\
\bottomrule
\end{tabular}
}

    \caption{Annotations for all 28 papers across the 24 initial contextual aspects (see \Cref{app:mapping-features}). We score each paper’s discussion of these as 2 (major argument), 1 (minor argument), and 0 (no mention).}
    \label{app:tab-lit-all-features}
\end{table*}

\subsection{Mapping of initial aspects}\label{app:mapping-features}
Our initial literature review yielded 24 aspects listed below that were mapped ($\rightarrow{}$) to the principles and features:
\begin{enumerate}
    \item Level of access auditors had $\rightarrow$ Access and Resources
    \item How did they work with the model developers $\rightarrow$ Integrity
    \item Were they trained by the developers to evaluate the model $\rightarrow$ Who are the auditors
    \item How much compute given to auditors, which manufacturers/chip model $\rightarrow$ Access and Resources
    \item Expertise / background of people involved $\rightarrow$ Who are the auditors
    \item What was their initial goal for the evaluation $\rightarrow$ What is evaluated
    \item What were their documentation requirements $\rightarrow$ Review and communication
    \item What was the contract structure / How were they paid/rewarded $\rightarrow$ Integrity
    \item What human input was used / who were they $\rightarrow$ How is it evaluated
    \item Peer review process / QA testing $\rightarrow$ Review and communication
    \item How does capability or concept translate to benchmark task $\rightarrow$ How is it evaluated
    \item Involvement of domain experts during design $\rightarrow$ Who are the auditors
    \item Feedback channel (openness + how to contact) $\rightarrow$ Review and communication
    \item How should scores be interpreted or used $\rightarrow$ How is it evaluated
    \item Assumptions of normative properties are documented $\rightarrow$ Assumptions
    \item Non-normative assumptions justified $\rightarrow$ Assumptions
    \item Evals as a process should consider eval's limitations (not that evals are limited) $\rightarrow$ Limitations
    \item Interdisciplinary teams / including multiple perspectives $\rightarrow$ Who are the auditors
    \item What is the technical goal of the model being evaluated $\rightarrow$ What is evaluated
    \item What are the principles behind the development of the model $\rightarrow$ What is evaluated
    \item Consider context in which the AI system will be used (e.g. value hierarchies, diverse populations) $\rightarrow$ Assumptions
    \item Justification of evaluation methods used / using qualitative evaluations as well $\rightarrow$ Justification
    \item Timely, continuous evals $\rightarrow$ What is evaluated
    \item Interpretable, easy to read, accessible to a wide audience $\rightarrow$ Review and communication
\end{enumerate}

\section{Audit cards} \label{app:audit_cards}

\subsection{Details and justifications of features}\label{app:details-features}
This section provides an overview of audit card features, including a concise summary of what to report and why this contextual information matters. 
A detailed list of aspects for each feature can be found in the checklist in \Cref{app:eval-card-example}.

\textbf{Who are the auditors?}

\textbf{Summary:} Documents auditor qualifications through domain expertise, experience, technical familiarity, formal credentials, stakeholder perspectives, and social positionality.

\textbf{Why it matters:} Transparency about qualifications establishes credibility while revealing potential biases. Expertise levels directly affect the ability to assess specific systems, while diverse backgrounds help identify evaluation blind spots. Without this information, readers cannot judge whether auditors possessed the necessary skills to thoroughly evaluate the system or what perspectives might be underrepresented in the analysis.

\textbf{Relevance across contexts:} Basic expertise documentation is essential for all evaluation types. Red-teaming specifically requires specialized adversarial thinking skills and security expertise. In user studies, auditor social positionality becomes particularly important as it may influence how human-AI interactions are interpreted.

\textbf{What is evaluated?}

\textbf{Summary:} Defines evaluation scope, goals, and approach by specifying system version, components, scaffolding used, threat models addressed, and whether capabilities or propensities were evaluated. Includes continuous monitoring plans and obsolescence criteria.

\textbf{Why it matters:} Clear boundaries prevent misinterpretation of results. Understanding threats, contexts, and capability vs. propensity assessment helps readers gauge relevance to their specific concerns. Without specific scope information, readers might incorrectly assume the evaluation covers more than it does, while obsolescence criteria help prevent outdated evaluations from being applied to newer system versions.

\textbf{Relevance across contexts:} Evaluation scope and goal specification are critical across all evaluation types. Benchmarks particularly require clear documentation of obsolescence criteria as AI capabilities rapidly evolve. Safety assessments benefit from explicit threat models tailored to deployment context. Human-centered evaluations need clear articulation of intended user groups and use cases, while deployed systems require specified intervals and conditions for re-evaluation.

\textbf{How is it evaluated?}

\textbf{Summary:} Describes methodology, implementation details, and result interpretation guidance. Explains connections between evaluation methods and capabilities, justifies capability-to-goal translations, and outlines finding limitations. Reporting on this feature can be very extensive, as the specific details for reporting depend on the exact evaluation procedure used. We refer to existing literature on best practices for reporting on evaluation procedures, such as \citet{reuelBetterBenchAssessingAI2024, WeNeedScience}.

\textbf{Why it matters:} Methodology directly impacts result validity and reliability. Understanding how capabilities become measurable criteria and how scores translate to real-world performance prevents misapplication of evaluation outcomes. Without methodological transparency, readers cannot assess whether the evaluation approach actually measures what it claims to measure or understand the confidence level appropriate for various findings.

\textbf{Relevance across contexts:} A description of methodology is essential for all evaluation types. Benchmarks specifically require detailed score interpretation guidance to prevent misapplication of numerical results. Propensity evaluations need robust justification for capability-to-goal translations since they often use proxy measurements. Novel system types need clearer justification for how evaluation approaches measure intended capabilities.

\textbf{Access and resources}

\textbf{Summary:} Details auditor system access level (black/gray/white-box, with/without safeguards), provided documentation, and available computational, time, and financial resources.

\textbf{Why it matters:} Resource constraints and access limitations impact evaluation thoroughness and depth. Limited system access or insufficient resources may prevent auditors from discovering certain vulnerabilities or testing edge cases. Transparency about these constraints helps readers understand which areas received less scrutiny and what types of issues the evaluation might have missed due to practical limitations.

\textbf{Relevance across contexts:} Basic access level information is relevant to all evaluations. Red-teaming findings can differ dramatically between black-box and white-box conditions. Large-scale benchmarks particularly require computational resource documentation.

\textbf{Integrity}

\textbf{Summary:} Addresses auditor selection process, potential ``opinion shopping,'' conflicts of interest, compensation structures, contract terms, and NDAs. Documents measures ensuring diverse perspectives.

\textbf{Why it matters:} Selection processes and incentive structures reveal potential influences on outcomes. Auditors with financial ties to the evaluated organization may face pressure to produce favorable results, while restrictive NDAs might limit disclosure of critical findings. Transparent disclosure of relationships and contractual limitations builds trust in findings and allows readers to identify potential sources of bias.

\textbf{Relevance across contexts}: Integrity documentation and conflict of interest disclosure is universally important across all evaluation contexts. Selection process transparency becomes critical for high-stakes evaluations where findings may significantly impact deployment decisions. Non-disclosure agreement details matter most when evaluations involve proprietary information that might limit the disclosure of certain findings. Disclosure of prior relationships between auditors and developers matters more for external evaluations to establish credibility and independence.

\textbf{Review and communication}

\textbf{Summary:} Outlines evaluation review process, feedback channels, executive summary quality, key term definitions, and result distribution. Focuses on unambiguous key findings and consistent terminology.

\textbf{Why it matters:} Robust review strengthens findings' reliability by catching errors and biases. Clear feedback channels enable continuous improvement when new information emerges or vulnerabilities are discovered after publication. Well-defined terminology ensures consistent understanding across different stakeholders, preventing miscommunication about technical concepts and evaluation outcomes that could lead to inappropriate system deployment decisions.

\textbf{Relevance across contexts:} Robust review processes are essential across all evaluation types to ensure reliability and catch potential errors or biases. Interdisciplinary evaluations require clear definitions of domain-specific terminology. Security-sensitive applications need explicit documentation of which findings are shared with which audiences.

\subsection{Checklist for audit card} \label{app:eval-card-example}
This checklist provides a framework for documenting contextual information about AI system audits. Questions are prioritized with the most important appearing first in each section, though not all questions in a section will be relevant to every evaluation. The checklist serves as a flexible guide rather than a rigid requirement. More comprehensive information generally enhances transparency. Through use of the audit card checklist, it is easier to document evaluation contexts, limitations, and strengths to facilitate informed discussions about AI system capabilities and risks.

\renewcommand{\labelitemiii}{$\bullet$}

\begin{itemize}[leftmargin=*]
    \checkbox \textbf{Who are the auditors}
    \begin{itemize}
        \checkbox \textbf{Expertise:}
        \begin{itemize}
            \item What specific domain expertise (e.g. cybersecurity knowledge) do the auditors possess that is relevant to this evaluation?
            \item What relevant experience do the auditors have in evaluating similar systems?
            \item What is the auditors' familiarity with the specific technology being evaluated?
            \item Do the auditors hold official accreditations or certifications relevant to this evaluation?
            \item Have the auditors undergone any training specific to this evaluation (internal or external)?
        \end{itemize}
        
        \checkbox \textbf{Background:}
        \begin{itemize}
            \item What stakeholder perspectives do the auditors represent or have experience with?
            \item What is the social positionality of the auditors (aggregated demographic information, cultural background)?
        \end{itemize}
    \end{itemize}

    \checkbox \textbf{What is evaluated}
    \begin{itemize}
        \checkbox \textbf{Scope:}
        \begin{itemize}
            \item Which specific version or iteration of the system is being evaluated?
            \item What specific components of the system are included in the evaluation?
            \item What aspects of the system were explicitly excluded from the evaluation scope?
            \item Does the evaluation cover the full system or only specific functionalities?
            \item What scaffolding (e.g. chain-of-thought prompting, few-shot demonstrations) or additional tools (e.g. web or command line access) were used during the evaluation?
        \end{itemize}
        
        \checkbox \textbf{Goal:}
        \begin{itemize}
            \item What threat models or risk scenarios is the evaluation addressing?
            \item What specific harms or misuses is the evaluation attempting to identify?
            \item What are the specific contexts (e.g. user groups and use cases) in which the AI system will be deployed?
            \item What are the performance expectations in the intended contexts?
        \end{itemize}
        
        \checkbox \textbf{Type of evaluation:}
        \begin{itemize}
            \item Is the evaluation assessing capabilities, propensities, or risks?
            \item What form of evaluation methodology is being used (benchmark, red-teaming, human uplift studies, or user studies, etc.)?
        \end{itemize}
        
        \checkbox \textbf{Continuous evaluations:}
        \begin{itemize}
            \item Will the system be subject to ongoing monitoring after this evaluation?
            \item At what intervals will follow-up evaluations occur?
        \end{itemize}
        
        \checkbox \textbf{Obsolescence criteria:}
        \begin{itemize}
            \item What changes (system updates, deployment context, discoveries about similar systems, or technological advances) would render this evaluation obsolete?
            \item How long is this evaluation expected to remain relevant without updates?
        \end{itemize}
    \end{itemize}

    \checkbox \textbf{How is it evaluated}
    \begin{itemize}
        \checkbox \textbf{Description of evaluation setup:}
        \begin{itemize}
            \item What general evaluation methodology was employed?
            \item What were the key components of the evaluation design?
            \item What evaluation framework or established protocol was followed (if applicable)?
        \end{itemize}
        
        \checkbox \textbf{Justification of capability-to-goal translation:}
        \begin{itemize}
            \item How does the evaluation approach align with the capabilities or risks being assessed?
            \item How do the evaluation procedures relate to the system's intended use contexts?
            \item What is the rationale for how the chosen methods measure the targeted capabilities?
        \end{itemize}
        
        \checkbox \textbf{Interpretation of scores:}
        \begin{itemize}
            \item How should the evaluation results be interpreted?
            \item What limitations should be considered when interpreting the findings?
            \item What do different performance levels indicate about the system?
        \end{itemize}
    \end{itemize}

    \checkbox \textbf{Access and resources}
    \begin{itemize}
        \checkbox \textbf{Level of access:}
        \begin{itemize}
            \item Did the auditor have black-box, gray-box, or white-box access?
            \item What specific system documentation was provided to auditors, including access to system code, architecture, design documents, training data, or previous evaluation results?
            \item Were any aspects of the system explicitly restricted from auditor access?
        \end{itemize}
        
        \checkbox \textbf{Resources available:}
        \begin{itemize}
            \item What computational resources were allocated to the evaluation?
            \item How much time was allocated for the full evaluation cycle?
            \item Was the evaluation timeline sufficient for comprehensive testing?
            \item How many auditors were involved and what was their time commitment?
            \item What specialized tools or software were available to the auditors?
            \item Were resources sufficient to conduct all planned evaluation activities?
            \item What resource constraints limited the scope or depth of the evaluation?
        \end{itemize}
    \end{itemize}

    \checkbox \textbf{Integrity}
    \begin{itemize}
        \checkbox \textbf{Selection process:}
        \begin{itemize}
            \item How were auditors identified and recruited for this evaluation?
            \item What criteria were used to select auditors?
            \item What measures were taken to ensure diversity of perspectives among auditors?
            \item Was there an open call or nomination process for auditors?
            \item Were independent third parties involved in the selection process?
        \end{itemize}
        
        \checkbox \textbf{Conflicts of interest:}
        \begin{itemize}
            \item What relationships exist between auditors and the organization being evaluated?
            \item Have auditors previously received compensation from the evaluated organization?
            \item Do auditors have financial interests or competing interests related to the evaluated system?
            \item What measures were taken to mitigate potential conflicts of interest?
            \item Were potential conflicts of interest publicly disclosed?
        \end{itemize}
        
        \checkbox \textbf{Compensation and incentive structures:}
        \begin{itemize}
            \item What was the compensation structure for auditors (fixed fee, hourly, etc.)?
            \item What contractual limitations were placed on auditors (non-disclosure agreements, etc.)?
            \item Were there incentives for identifying system flaws or vulnerabilities?
            \item Were auditors employed internally or contracted externally?
            \item Were there any performance-based incentives that might bias results?
        \end{itemize}
    \end{itemize}

    \checkbox \textbf{Review and communication}
    \begin{itemize}
        \checkbox \textbf{Review:}
        \begin{itemize}
            \item What aspects of the evaluation (e.g. methodology, report, findings) were subjected to review?
            \item Who reviewed these aspects, and what was their relationship to the auditors and system developers?
            \item What expertise did the reviewers bring to the process?
            \item How were reviewer disagreements resolved and what changes resulted from the review process?
        \end{itemize}
        
        \checkbox \textbf{Feedback channels:}
        \begin{itemize}
            \item What specific mechanisms exist for providing feedback on the audit?
            \item How can new information be submitted for consideration after publication?
            \item Who should be contacted with questions or concerns about the evaluation?
            \item Is there a plan for addressing follow-up questions about the results?
        \end{itemize}
        
        \checkbox \textbf{Executive summary:}
        \begin{itemize}
            \item Does the summary clearly explain what was evaluated, how, and what was found?
            \item Are limitations of the evaluation explicitly stated?
            \item Is the summary accessible to non-technical audiences?
        \end{itemize}
        
        \checkbox \textbf{Definitions of key terms:}
        \begin{itemize}
            \item Are technical terms and evaluation metrics clearly defined?
            \item Is terminology consistent with industry standards?
        \end{itemize}
        
        \checkbox \textbf{Access to results:}
        \begin{itemize}
            \item Who will receive which parts of the evaluation findings?
            \item How and when will results be published or distributed?
            \item What information, if any, will be redacted from public versions?
        \end{itemize}
    \end{itemize}
\end{itemize}

\section{Selection of governance documents} \label{app:gov-selection}
We chose the relevant governance documents to include in the analysis by considering the following three factors:
\begin{enumerate}
    \item Major jurisdiction: broadly selected by geopolitical and economic centrality, especially in the AI industry \citep{stanfordhaistaffGlobalAIPower2024,bianzinoArtificialIntelligenceAI2023}.
    \item Authority of the institution: within each jurisdiction, we chose government and non-government institutions. METR was the exception because their evaluation reports have been highly influential in the field, and often involve co-authorship from researchers in scaling labs and AISIs.
    \item Direct influence on model evaluations. For model developers, these tended to be their Responsible Scaling Policies or Frontier Safety Frameworks (see \citealp{metrFrontierAISafety}). For government bodies, we selected documents that tracked with more scale and influence (i.e. more national/federal than local, or more binding than not). The EU Code of Practice is a unique outlier in not being strictly binding but in effect being as influential as it enables presumption of conformity with the Act (see \citet{TypesLegislationEuropean} and \citet{eumonitorLegalInstruments}).
\end{enumerate}

To enable the governance documents to fit \Cref{tab:governance}, we abbreviated the names of the documents. Here, we present their official names.

\begin{itemize}
    \item \textbf{Official regulations}
    \begin{itemize}
        \item Brazil: We analyzed the Federal Senate Bill No. 2338/2023 \citep{BrazilAIAct2023}. It was approved by Senate in 2024, but still pending approval by the House of Representatives and President to become law.  
        \item China: Governance of AI models and systems is detailed across various documents. This analysis accounts for: Security Specification for Generative AI Pre-training and Fine-tuning Data \citep{chinaGenAI2025}, Draft AI Law, translated with scholars comments \citep{murphyArtificialIntelligenceLaw}; Internet Information Service Algorithmic Recommendation Management Provisions with effect from 2022 \citep{creemersTranslationInternetInformation2022}; Measures for the Management of Generative Artificial Intelligence Services (Draft for Comment) 2023 \citep{huangTranslationMeasuresManagement2023}; China 2022 rules for deep synthesis \citep{guChinaReleasedNew2023}; and China (Draft Measure on Ethical Review).
        \item EU: We included the keystone AI Act \citep{euAIAct}; with its accompanying Code of Practice (``CoP'') \citep{EUGPAI2025}. The EU AI Act needs to be read in conjunction with the Code of Practice. The Code of Practice is a unique type of document which, while technically not legally binding, enables presumption of conformity with the EU AI Act. In effect, due to the legal uncertainty of not knowing what alternatives would comply, it is likely to be followed by those wishing to be compliant with the EU AI Act, resulting in effectively a binding effect until ``harmonized standards'' are published. 
        \item Singapore: We analyzed the Model AI Governance Framework, which is not technically binding but authoritative guidance. Model AI Governance Framework for Generative AI \citep{SGgov2024}.
        \item South Korea: We analyzed the ``Basic Act on the Development of Artificial Intelligence and Establishment of Foundation for Trust'', which comes into force Jan 2026. Governance of AI models and systems appears to be done by a keystone AI Act with effect from 2026 \citep{SouthKoreaAIAct2024}. 
        \item US: Given the lack of a US federal-wide AI regulation, we examined the Trump Administration's AI Action Plan issued in July 2025. This does not have the status of binding law but is the most authoritative signal at the federal level as to how the US government might think about governing model evaluations and risk assessments \citep{USActionPlan2025}. 
    \end{itemize}
    
    \item \textbf{Other norms}
    \begin{itemize}
        \item Bletchley: The Bletchley Declaration by Countries Attending the AI Safety Summit, 1-2 November 2023 \citep{bletchley2023} 
        \item Int'l Report: International AI Safety Report \citep{intlaisafety2025} 
        \item Japan AISI: Guide to Evaluation Perspectives \citep{japanaisi2024} 
        \item METR: Key Components of an RSP \citep{metr2024} 
        \item NIST AI RMF: AI Risk Management Framework \citep{nist2024} 
        \item Paris: Statement on Inclusive and Sustainable Artificial Intelligence for People and the Planet \citep{paris2025} 
        \item Seoul: Seoul Declaration for Safe, Innovative and Inclusive AI \citep{seoul2024} 
        \item UK AISI: AI Safety Institute Approach to Evaluations \citep{ukaisi2024} 
    \end{itemize}
    
    \item \textbf{AI company policies}
    \begin{itemize}
       \item Anthropic: Responsible Scaling Policy \citep{anthropicResponsibleScalingPolicy2025} 
        \item Google DeepMind: Frontier Safety Framework v2 \citep{FrontierSafetyFramework} 
        \item Meta: Frontier AI Framework v1.1 \citep{metaFrontierAIFramework} 
        \item Microsoft: Frontier Governance Framework \citep{microsoftFrontierGovernanceFramework2025} 
        \item OpenAI: Preparedness Framework Beta \citep{openaiPreparednessFramework2025} 
        \item xAI: Draft Risk Management Framework \citep{XAIRiskManagement} 
    \end{itemize}
\end{itemize}

\section{Stakeholder interviews}\label{app:interviews}
\subsection{Detailed Methodology}
\textbf{We engaged with stakeholders through interviews in two stages}.
During literature review and feature selection (see \Cref{sec:eval-cards}), we engaged in a dozen off-the-record discussions with academics, evaluation report writers, and policy researchers. After completing the audit card framework, we engaged in the ten semi-structured interviews with stakeholder experts (see \Cref{sec:interviews}). Our protocol focused on current practices, perceived gaps, and contextual needs.

\textbf{During both stakeholder engagement stages, the feedback shaped our audit card design}. For instance, we distinguished compensation from conflicts of interest (``Integrity"), or condensed features within ``Review and feedback". Interviewees also generally agreed with the high-level value of this audit card proposal, but some expressed doubts about implementation, resulting in the addition of Appendix C. 

\textbf{The interview process was fully based on informed consent.} We implemented a comprehensive consent process, obtaining explicit agreement on recording, attribution levels, and data usage. Participants chose their preferred anonymity level, and all quotes were verified before inclusion. Participants were not compensated for the 30-minute voluntary interviews. 

Our procedure for reaching out to interviewees was as follows:
\begin{quote}
As part of the [anonymized for review], we are working on a research project on reporting contextual aspects of the AI evaluations process. The goal is to improve the quality of evaluations through clearer norms on responsible reporting. Here is a more detailed project summary, and below you find the types of information an extensive literature survey has shown to be relevant to evaluations reporting. 

As part of this project, we are now looking to get the input from diverse stakeholders in the evaluations process. This could either be in writing or during a short 30-minute call. We would be grateful if you could answer a few questions, which, depending on your exact responsibilities as [role of interviewee], would be some subset of the following:

[Questions as shown in \Cref{app:interview-guide}.]

Any contribution of yours will be appropriately cited following academic standards. Please let me know if you have any preferences regarding direct quotation or paraphrasing. We are happy to adjust attribution to reflect your preferred level of visibility, whether by name, role at [interviewee organisation], or role more generally. We are happy to share any citations or attributions with you before publication to ensure accuracy and alignment with your preferences.
\end{quote}

Finally, we followed up with interviewees to confirm the level of attribution:
\begin{quote}
We'd be very grateful if you could let us know by Wednesday EOD if you are alright with how this information is being presented. If it's not acceptable to you, we'd be grateful to understand why and how we can better frame it for better accuracy or anonymity. Any changes to the final draft will be in the direction of less detail/identifiability. 

This is how your identity is described: [description of interviewee as per \Cref{tab:interviewee-table}.]  

This is how your interview is being attributed (you are [$P_x$], highlighted is where you're cited alone, $P_y-P_z$ indicate other interviews being referenced together with yours): 

[Selection from \Cref{sec:interviews} that references the interviewee.]

\end{quote}

\subsection{Interviewee details}
See \Cref{tab:interviewee-table} for details on the background of interviewees.
\begin{table*}[h!]
    \centering
    \begin{tabular}{|l|l|l|}
    \hline
    \setlength{\tabcolsep}{3mm}
    \textbf{Participant} & \textbf{Organization} & \textbf{Role} \\
    \hline
    P1 & Third-party evaluator& Evaluation designer, developer \& writer\\
    \hline
    P2 & [anonymized] & [anonymized] \\
    \hline
    P3 & Scaling lab& Evaluation developer\\
    \hline
    P4 & Third-party evaluator& Evaluation developer\\
    \hline
    P5 & Third-party evaluator& Evaluation developer \& writer\\
    \hline
    P6 & [anonymized] & Evaluation policy researcher\\
    \hline
    P7 & Third-party& Evaluation policy researcher\\
    \hline
    P8 & Government& Evaluation policymaker \& writer\\
    \hline
    P9 & Third-party evaluator& Evaluation development operations support\\
    \hline
    P10 & [anonymized] & [anonymized] \\
    \hline
    \end{tabular}
    \caption{Overview of the stakeholders interviewed for the semi-structured interviews. The organization and role has been reported up to the level of detail interviewees were comfortable with. Note: Participants P2 and P10 were excluded from the final analysis as their interviews were conducted off the record and attributions could not be confirmed in time for publication.}
    \label{tab:interviewee-table}
\end{table*}

\subsection{Interview guide}\label{app:interview-guide}
The following is our interview guide and set of questions we used for the 30-minute semi-structured interview with each interviewee.

\noindent\textbf{Introductions [3 mins]}
\begin{itemize}
  \item Introduce our project: We defined what audit cards are, what we did and did not include in that scope
  \begin{itemize}
    \item Audit cards are a way to standardize reporting of AI evaluation process, with the goal of contextualizing what evaluations can do and transparency about the process
    \item We state examples of what we include: why these metrics, who are the auditors, resources available (compute, time) during eval, how they're selected and potential COI
    \item We disambiguate them from examples we don't include: specific metrics, technical implications, elicitation techniques;
  \end{itemize}
  \item Outline process for interview: We walked interviewees through what to expect of the interview
  \begin{itemize}
    \item Contribution: Used to color in details about evaluations process from stakeholders.\\
    \item Privacy and permissions: We checked beforehand what level of recording, attribution, and other process checks they would be comfortable with. We confirmed nothing they said to us will be shared to others without it being run past them and them confirming in writing.
  \end{itemize}
\end{itemize}

\noindent\textbf{General questions to ask everyone [10 mins max]}
\begin{enumerate}
  \item Describe the work you do in relation to evaluations, try to be specific about your responsibilities, identities of people/orgs you relate to, your role in the evaluations field
  \item Who is the evaluation report you [build/analyze/use] for? Who is the target audience? E.g. developers (internal, external), research, governance, or also the general public.
  \item What part of the evaluation process which, if not done well/reported well, would make you doubt the quality of the eval?
  \item How do imagine evaluation best practices best becoming reality? Do you see codification---whether in law or Industry norm---as a good/valid way to quality control?
  \item In your view, what are the biggest challenge preventing evaluations from being better? More useful for improving safety, or whatever their key goal is?
\end{enumerate}

\noindent\textbf{Questions to ask specific categories of people [12 mins]}

\noindent\textbf{Evaluation developer}
\begin{enumerate}
  \setcounter{enumi}{5}
  \item Do you consider the limitations and assumptions of your evaluations process? Where does that thinking get captured (if it does)?
  \item What is the background/expertise of auditors—does it depend on the type of evaluation or something else, and who makes that decision?
  \item Can you give us examples of evaluations where the evaluations process differed and whether it affected the quality of eval?
  \item Do you follow any internal best practices?
  \item What is the current training process both internally (with evaluation developers, human baseliners) and with your engagement partner?
\end{enumerate}

\noindent\textbf{Evaluation report writer}
\begin{enumerate}
  \setcounter{enumi}{10}
  \item What is the most important information to share in the evaluation report? How do you decide that?
  \begin{enumerate}
    \item Can you walk through two situations in which you decided differently (i.e. type of info/detail level to publish and why)?
    \item What's the right balance of transparency in reporting evaluations, and what risks surround the achieving of that?
  \end{enumerate}
  \item From our skim of evaluation reports, we found these tend to be underspecified/not specified—why do you think that might be? e.g. assumptions, auditors' background, integrity and resources of process, peer review, commitments to take action based on evaluation results, and state criteria that make evaluation obsolete
\end{enumerate}

\noindent\textbf{Evaluation development overseer/manager \& translate it to policy people}
\begin{enumerate}
  \setcounter{enumi}{12}
  \item Do you consider the limitations and assumptions of your evaluations? Where does that thinking get captured (if it does)?
  \begin{enumerate}
    \item e.g. specific downstream application context on which the benchmark is contingent;
  \end{enumerate}
  \item How do you select the right auditors (external / internal)? Walk through a recent eval's selection?
  \item How much autonomy does your organization get in deciding access/resources (e.g. time/compute) for evaluations [of a lab's models]? What does this depend on?
  \begin{enumerate}
    \item Do you include different stakeholders in the evaluations process—who are they, and how are they engaged?
  \end{enumerate}
  \item What is the current training process both internally (with evaluation developers, human baseliners) and with your engagement partner?
  \item What is the right balance of flexibility and specificity in regulation? E.g. what makes an auditor 'qualified' (background/training), conflict of interest, funding/contract structures
  \item If you wanted to, how easy would it be for you to be misleading when communicating evaluation results? Why?
\end{enumerate}

\noindent\textbf{Evaluation operations support} (decisions getting access to resources that evaluation developers need)
\begin{enumerate}
  \setcounter{enumi}{18}
  \item What resources do you usually require for an eval, and what influences that? Do you document resource requirements, and where do you share this?
  \item What are the conditions of you being able to access them? E.g. Recent Cyber CBRN Agent evaluations report used blue/red anonymized models at UK AISI
  \item Could you describe the different engagement processes, and why some have been easier than others?
\end{enumerate}

\noindent\textbf{Evaluation used for recommendations} - policy researcher/think tank, funders, lobbying, advocacy
\begin{enumerate}
  \setcounter{enumi}{21}
  \item If you have had to use an evaluation report (could be your org or another org.) to make a recommendation, what features of the report have helped you to do so?
  \item Do you have examples of/from evaluation reports you found good and/or bad, and why?
  \item How much do you trust evaluation results depending on the funding source (e.g. internal different labs OpenAI, external: Apollo, AISI)
\end{enumerate}

\noindent\textbf{AI safety communicators}
\begin{enumerate}
  \setcounter{enumi}{24}
  \item How do you understand evaluations (as distinct from audits) and what type of evaluation reporting/communication are you aware of?
  \item What insights are people most often seeking from evaluations, and how are evaluations (not) meeting that need?
  \item Re target audience question: Why do you think [previous answer] is the relevant audience? Other relevant audience segments, how does the messaging change? How can evaluations most strongly communicate their goal?
\end{enumerate}

\end{appendices}

\end{document}